\documentclass[aps,pre,reprint,groupedaddress]{revtex4-2}
\usepackage{amsmath,amssymb,amsthm,dcolumn}
\usepackage{graphicx}
\usepackage{xcolor}
\usepackage{newtxtext}
\usepackage[slantedGreek,cmintegrals]{newtxmath}
\usepackage{hyperref}
\hypersetup{colorlinks,linkcolor={blue!90!black},citecolor={blue!90!black},urlcolor={blue!90!black}}
\usepackage{stmaryrd}
\usepackage{enumitem}
\usepackage{accents}
\usepackage[compact]{titlesec}
\DeclareMathAlphabet{\mathcal}{OMS}{cmsy}{m}{n}
\definecolor{linkcolor}{HTML}{0000ff}
\hypersetup{colorlinks,linkcolor={linkcolor},citecolor={linkcolor},urlcolor={linkcolor}}
\renewcommand{\pi}{\uppi}
\begin{document}
\title{Euler buckling on curved surfaces}

\author{Shiheng Zhao}
\affiliation{Max Planck Institute for the Physics of Complex Systems, N\"othnitzer Straße 38, 01187 Dresden, Germany}
\affiliation{Center for Systems Biology Dresden, Pfotenhauerstraße 108, 01307 Dresden, Germany}
\affiliation{\mbox{Max Planck Institute of Molecular Cell Biology and Genetics, Pfotenhauerstraße 108, 01307 Dresden, Germany}}
\author{Pierre A. Haas}
\email{haas@pks.mpg.de}
\affiliation{Max Planck Institute for the Physics of Complex Systems, N\"othnitzer Straße 38, 01187 Dresden, Germany}
\affiliation{Center for Systems Biology Dresden, Pfotenhauerstraße 108, 01307 Dresden, Germany}
\affiliation{\mbox{Max Planck Institute of Molecular Cell Biology and Genetics, Pfotenhauerstraße 108, 01307 Dresden, Germany}}

\renewcommand{\vec}[1]{\boldsymbol{#1}}
\renewcommand{\eqref}[1]{Eq.~(\ref{#1})}
\newcommand{\eqsref}[1]{Eqs.~(\ref{#1})}
\newcommand{\neqref}[1]{(\ref{#1})}
\newcommand{\figref}[2]{(Fig.~\hyperref[#1]{\ref*{#1}#2})}
\newcommand{\textfigref}[2]{Fig.~\hyperref[#1]{\ref*{#1}#2}}
\newcommand{\MM}{\hyperref[mm]{\emph{Materials and Methods}}}

\begin{abstract}
Euler buckling epitomises mechanical instabilities: An inextensible straight elastic line buckles under compression when the compressive force reaches a critical value $F_\ast>0$. Here, we extend this classical, planar instability to the buckling under compression of an inextensible relaxed elastic line on a curved surface. By weakly nonlinear analysis of an asymptotically short elastic line, we reveal that the buckling bifurcation changes fundamentally: The critical force for the lowest buckling mode is $F_\ast=0$ and higher buckling modes disconnect from the undeformed branch to connect in pairs. Solving the buckling problem numerically, we additionally find a new post-buckling instability: A long elastic line on a curved surface snaps through under sufficient compression. Our results thus set the foundations for understanding the buckling instabilities on curved surfaces that pervade the emergence of shape in biology.
\end{abstract}

\maketitle

\section{\uppercase{Introduction}}

In 1744, \textsc{Euler}~\cite{euler1744} described the paradigm of a mechanical instability~\cite{timoshenko}: An elastic rod of length $\ell$ compressed by a force $F$ applied to its ends buckles when~$F$ reaches a critical load $F_\ast\propto \ell^{-2}$. Two decades later, \textsc{Lagrange}~\cite{lagrange1770} analysed the higher buckling modes that become possible as $F$ is increased further, but the full nonlinear description of this fundamental instability~\cite{goriely2008} is a culmination of two more centuries of work that saw the development of Euler's model of elasticity into the modern field theory of solid mechanics~\cite{ogden}. 

Much more recently, the importance of related mechanical instabilities for the emergence of biological shape has begun to be appreciated~\cite{nelson16}. For example, a buckling instability is believed to cause the twisting phenotypes of germband extension in \emph{Drosophila} mutants~\cite{celia2023} or the symmetry breaking of its primordial hindgut~\cite{alber24}. Unlike classical Euler buckling however, this instability occurs within the curved surface of the \emph{Drosophila} embryonic tissue. This puts the finger on a glaring gap in our understanding of mechanical instabilities: Even the most basic buckling problems on general surfaces remain wide open.

Here, we therefore analyse a minimal buckling instability within a curved surface: the Euler buckling of a compressed inextensible elastic line of length $\ell$ on a smooth surface $z=h(x,y)$ embedded in three-dimensional Euclidean space with Cartesian coordinates $(x,y,z)$. The buckled shape minimises the bending energy of the elastic line subject to the imposed compression and the constraint of inextensibility. We parameterise the elastic line by its arclength $s$, so that a point on it has coordinates $\smash{\bigl(x(s),y(s),z(s)\bigr)}$, with $\smash{z(s)=h\bigl(x(s),y(s)\bigr)}$. The Lagrangian of the problem is
\begin{align}
\mathcal{L}=\dfrac{1}{2}\int_0^\ell{\kappa(s)^2\,\mathrm{d}s}-\int_0^\ell{\lambda(s)\bigl[\alpha(s)^2-1\bigr]\,\mathrm{d}s},\label{eq:L}
\end{align}
where $\kappa(s)^2=x''(s)^2+y''(s)^2+z''(s)^2$ is the squared curvature of the elastic line and $\alpha(s)^2=x'(s)^2+y'(s)^2+z'(s)^2$ is its stretch squared, with dashes denoting differentiation with respect to $s$. The Lagrange multiplier function $\lambda(s)$ imposes inextensibility, $\alpha(s)\equiv1$.

To define our buckling problem on this surface, we first paraphrase classical Euler buckling~\figref{fig1}{A}: In the plane, an elastic line with one clamped end and one free end relaxes into a straight line ($\kappa=0$). Clamping of the other end and compression, by a force $F$, along the straight, relaxed configuration by a relative amount $\delta$ leads to buckling. The ``up'' and ``down'' buckling modes are equivalent because of the symmetry of the plane. On a curved surface~\figref{fig1}{B} however, it is known that the relaxed elastic line need not be ``straight'', i.e., need not be a geodesic~\cite{Manning1987}. Clamping of the other end and compression along the relaxed shape will still lead to buckling, but the curvature of the surface will in general select either up or down buckling by breaking the symmetry of the two buckling modes.

To show that the curvature of the surface changes the buckling instability even more fundamentally, we will first solve the buckling of a short elastic line asymptotically, before studying long elastic lines numerically.

\section{\uppercase{Results}}
\subsection{Asymptotic buckling of a short elastic line} We solve the buckling of a short elastic line by asymptotic expansion for small $\ell$. Up to translation and rotation of the surface, its tangent plane at the origin coincides with the $(x,y)$ plane, and we may clamp one end of the elastic line there in the $x$-direction. Scaling lengths by $\ell$, we therefore take
\begin{align}
h(x,y)&=\ell\bigl(Ax^2+Bxy+Cy^2\bigr),\label{eq:hmain}
\end{align}
in which the parameters $A,B,C$ describe the surface. By solving the Euler--Lagrange equations of~\eqref{eq:L}, we compute the shape of the relaxed line order-by-order and hence solve the buckling problem (\MM).

\begin{figure}[t]
\centering
\includegraphics[width=\linewidth]{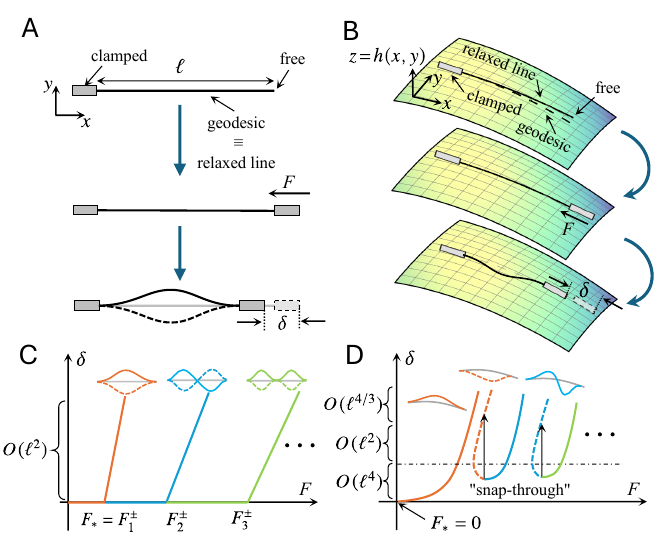}
\caption{Euler buckling in the plane and on a curved surface. (A)~Classical Euler buckling in the plane. An inextensible elastic line of length $\ell$, clamped at one end, relaxes into a straight, geodesic shape. Clamping of the other end and compression by a force $F$ along the straight shape leads to buckling. The relative compression is $\delta$. ``Up'' and ``down'' buckling (solid and dashed line in the bottom panel) is equivalent by symmetry. (B)~Euler buckling on a curved surface $z=h(x,y)$. The relaxed elastic line differs, in general, from the ``straight'' geodesic, and the curvature of the surface selects either ``up'' or ``down'' buckling. (C) Bifurcation diagram of classical Euler buckling for $\ell\ll 1$: Plot of relative compression $\delta$ against compressive force~$F$, from asymptotic analysis for $\delta=O(\ell^2)$. The first buckling mode (inset) appears at the critical force $F_\ast=F_1^\pm$ for ``up'' or ``down'' buckling. Higher buckling modes (insets) have higher critical forces $F_2^\pm,F_3^\pm,\dots$. (D)~Corresponding bifurcation diagram for a general curved surface: ``Up'' and ``down'' buckling modes disconnect and the asymptotic analysis at order $O(\ell^2)$ breaks down. Asymptotics for $\delta=O(\ell^4)$ show that $F_\ast=0$ and that higher modes connect in pairs. Asymptotics for $\delta=O(\ell^{4/3})$ show that they undergo a snap-through instability.}
\label{fig1}
\end{figure}

We first consider $\delta=O\bigl(\ell^2\bigr)$. Writing $\delta=\ell^2d_{(2)}$, we find that, in the lowest buckling mode,
\begin{align}
F &= \dfrac{4\pi^2}{\ell^2}\!-\!\dfrac{1}{\ell}\biggl(\dfrac{4|AB|\pi}{\sqrt{d_{(2)}}}\biggr)\!+\!\left[2d_{(2)}\pi^2\!+\!g(A,B,C)\right]+O(\ell),\label{eq:F2}
\end{align}
in which $g(A,B,C)=8A^2\left(\pi^2-1\right)+3B^2+12AC$. \eqref{eq:hmain} is the leading term in a Taylor expansion of a general surface near the origin; its corrections at order $\smash{O\bigl(\ell^2\bigr)}$ translate to corrections to \eqref{eq:F2} at order $O(1)$.

In the plane, this becomes $F=4\pi^2/\ell^2+2d_{(2)}\pi^2+O(\ell)$. By taking $d_{(2)}\to 0$, we recover the critical force for classical Euler buckling, ${F_\ast\sim4\pi^2/\ell^2}$. The stiffness of the buckled rod is $\mu_\ast=\partial F/\partial\delta\sim2\pi^2/\ell^2$. Higher buckling modes (\MM) are associated with higher forces, and complete the bifurcation diagram~\figref{fig1}{C}.

In general, we cannot however take $d_{(2)}\to 0$ in \eqref{eq:F2}, because it loses asymptoticity in this limit unless $AB=0$. The latter includes the cases of (i) a surface for which the clamping direction is a principal axis ($B=0$; e.g., a sphere) and (ii) a surface that is flat in the direction of clamping ($A=0$; e.g., clamping parallel to the axis of a cylinder). In these cases, $F_\ast\sim 4\pi^2/\ell^2+g(A,B,C)$ and $\mu_\ast\sim 2\pi^2/\ell^2$, i.e., the critical buckling force is merely shifted compared to the flat case, and the stiffness remains unchanged.

The case $AB\neq0$ is much more interesting: The relaxed elastic line is no longer symmetric under $y\mapsto-y$ (\MM). This breaks the symmetry of ``up'' and ``down'' buckling and turns the instability into a ``squeeze deformation'' with $F_\ast=0$~\figref{fig1}{D}. More precisely, the first two terms in \eqref{eq:F2} lose asymptoticity when $\smash{d_{(2)}=O\bigl(\ell^2\bigr)}$. We therefore make the ansatz $\smash{\delta=\ell^4d_{(4)}}$, with $d_{(4)}=O(1)$. We now find $\smash{F=\lambda_\ast^2/\ell^2+O\bigl(\ell^{-1}\bigr)}$, where $\lambda_\ast$ is determined implicitly by
\begin{align}
\dfrac{d_{(4)}}{(AB)^2}&=\frac{1}{90 \lambda_\ast\left(\lambda_\ast\sin{\lambda_\ast}+2\cos{\lambda_\ast}-2\right)^2}\left\{2\lambda_\ast\left[21 \lambda_\ast^2 +72\right.\right. \nonumber \\
&\quad+\left.\left(20\lambda_\ast^2-6\right)\cos{\lambda_\ast}+\left(4 \lambda_\ast^2-66\right)\cos{2\lambda_\ast}\right]\nonumber\\
&\quad-\!\left.3\!\left[\left(52\lambda_\ast^2\!+\!60\right)\sin{\lambda_\ast}\!+\!\left(19 \lambda_\ast^2\!-\!30\right)\sin{2\lambda_\ast}\right]\right\}\!.\label{eq:l}
\end{align}
In particular, $F\sim 175d_{(4)}/(AB\ell)^2$ for $d_{(4)}\ll 1$, which is consistent with $F_\ast=0$. This implies $\smash{\mu_\ast\sim 175\big/\bigl[(AB)^2\ell^6\bigr]\propto\ell^{-6}}$, so even the scaling of stiffness is modified from $\mu_\ast\propto\ell^{-2}$ of flat Euler buckling. For $\smash{d_{(4)}/(AB)^2\gtrapprox 0.21}$, \eqref{eq:l} has multiple additional solutions for $\lambda_\ast$ and hence $F$, which correspond to higher buckling modes~\figref{fig1}{D}. At local minima of $d_{(4)}$ as a function of $\lambda_\ast$ or $F$, pairs of different buckling modes merge. Importantly, these higher modes do not therefore connect to $\delta=0$, as they do in classical Euler buckling~\figref{fig1}{C,D}. 

Similarly, the second and third terms in \eqref{eq:F2} lose asymptoticity when $\smash{d_{(2)}=O\bigl(\ell^{-2/3}\bigr)}$, i.e., when $\smash{\delta=O\bigl(\ell^{4/3}\bigr)}$. With the ansatz $\smash{\delta=\ell^{4/3}d_{(4/3)}}$, where $d_{(4/3)}=O(1)$, we find
\begin{align}
F =  \dfrac{4\pi^2}{\ell^2} + \dfrac{1}{\ell^{2/3}}\biggl(2d_{(4/3)}\pi^2\mp\dfrac{4|AB|\pi}{\sqrt{d_{(4/3)}}}\biggr) + O(1),\label{eq:largeb}
\end{align}
for the lowest ($-$) and next ($+$) buckling modes, which are equal if $AB=0$. Thus, in the $-$ mode, $F$ continues to increase monotonically with compression, but $\partial F/\partial\delta<0$ on part of the $+$ branch, indicating a snap-through instability absent from classical Euler buckling \figref{fig1}{C,D}. Similar instabilities arise on higher branches (\MM).

\subsection{Intrinsic and extrinsic buckling} 
The bending energy in \eqref{eq:L} involves the total curvature $\kappa(s)^2=\kappa_{\text{g}}(s)^2+\kappa_{\text{n}}(s)^2$, where $\kappa_{\text{g}}$ and $\kappa_{\text{n}}$ are the geodesic and normal curvatures of the elastic line~\cite{docarmo}, so penalises both bending of the line \emph{in} the surface and its bending \emph{with} the surface. This is \emph{extrinsic buckling} because it involves $\kappa_{\text{n}}$, which is an extrinsic property of the surface. Instead, \emph{intrinsic buckling} replaces $\kappa\to\kappa_{\text{g}}$ in \eqref{eq:L}. In the latter case, the relaxed elastic line is always a geodesic, and, with $\delta=\ell^2d_{(2)}$, we find (\MM)
\begin{align}
F = \frac{4\pi^2}{\ell ^2} + 2d_{(2)}\pi^2 + K + 8A^2\pi^2 +O(\ell),\label{eq:Fint}
\end{align}
where $K = 4AC-B^2$ is the Gaussian curvature of the surface. Harking back to the \emph{Theorema egregium}~\cite{docarmo}, its appearance is unsurprising. However, this problem is not geometrically intrinsic: Because of the boundary clamping, the force on the ``right'' of the geodesic, given by \eqref{eq:Fint}, differs from that on its ``left'' by an amount $8A^2\pi^2+O(\ell)$ (\MM), which is not an intrinsic quantity. 

Meanwhile, $F_\ast\sim4\pi^2/\ell^2+K+8A^2\pi^2$ by taking ${d_{(2)}\to 0}$ in \eqref{eq:Fint}. Thus, even if $AB\neq 0$ and unlike extrinsic buckling, intrinsic buckling simply shifts the buckling threshold of classical, flat Euler buckling.

\begin{figure}
\centering\includegraphics[width=\linewidth]{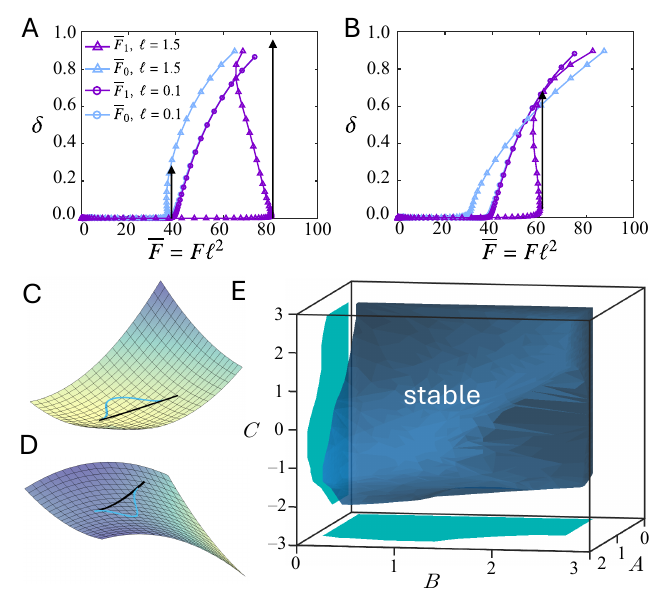}
\caption{Numerical Euler buckling on curved surfaces. (A)~Force-compression diagram for the lowest buckling mode on a surface of Gaussian curvature $K>0$ (${A=B=C=1}$), for short and long elastic lines ($\ell=0.1$, $\ell=1.5$). For $\ell=0.1$, the forces $F_0\neq F_1$ at either end increase monotonically with the compression $\delta$. For $\ell=1.5$, a snap-through instability arises ($\partial F_0/\partial \delta<0$, $\partial F_1/\partial \delta<0$; arrows). (B)~Analogous bifurcation diagram for a surface with $K<0$ (${-A=B=C=1}$). (C) Plot of the surface, relaxed elastic line (black), and buckled line (blue) in panel (A). (D)~Analogous plot for panel (B). (E) Phase diagram for this snap-through instability for an elastic line of fixed length $\ell=1$ in surface parameter space $(A,B,C)$, for $A,B\geqslant 0$. This is extended to all $A,B$ by the symmetries $(B,y)\mapsto-(B,y)$ and $(A,B,C)\mapsto-(A,B,C)$ of the problem.}\label{fig2}
\end{figure}

\subsection{Numerical buckling of a long elastic line} We now extend our asymptotic results by solving the Euler--Lagrange equations of \eqref{eq:L} numerically (\MM) for the surface defined by \eqref{eq:h}. For sufficiently short elastic lines, the forces $F_0\neq F_1$ at either end of the elastic line continue to increase monotonically with $\delta$~\figref{fig2}{A,B}. For longer elastic lines, however, we find an additional snap-through instability even of the lowest buckling mode, heralded by $\partial F_0/\partial\delta<0$ or $\partial F_1/\partial\delta<0$~\figref{fig2}{A,B}, that exists for surfaces both of positive~\figref{fig2}{A,C} and negative Gaussian curvature~\figref{fig2}{B,D}. The region of surface parameter space $(A,B,C)$ in which this new instability arises at fixed $\ell$ is, however, complex~\figref{fig2}{E}.

\section{\uppercase{Discussion}}
We have shown that even the simplest buckling instability \emph{within} curved surfaces displays rich additional behaviour compared to flat or ``simple'' curved surfaces. Similarly, mechanical instabilities \emph{of} curved surfaces driving brain gyrification are known to be affected by surface curvature~\cite{tallinen14,budday15}.

For our general surfaces, the effect of global curvature, relevant for elastic filaments on spherical bubbles~\cite{guven12,prasath21} for example, remains an open problem, as does, more mathematically, the effect on these instabilities of geometric singularities of the curved surface, such as folds or cusps.  

In this way, our work emphasises the importance of curvature differences for mechanical instabilities in morphogenesis morphogenesis. After all, cows are not spherical!

\appendix
\renewcommand{\thesection}{\textbf{APPENDIX \Alph{section}}}
\renewcommand{\theequation}{\Alph{section}\arabic{equation}}
\section{\uppercase{Materials \& Methods}}\label{mm}
\subsection{Boundary conditions} Let $L(s)=\kappa(s)^2/2-\lambda(s)[\alpha(s)^2-1]$ be the Lagrangian density in \eqref{eq:L}. Scaling lengths by $\ell$, the boundary terms in its variation are $\smash{\mathcal{B}=\left\llbracket\vec{f}(s)\cdot\delta\vec{r}(s)+\vec{m}(s)\cdot\delta\vec{r}'(s)\right\rrbracket_0^1}$, with
\begin{align}
&\vec{f}(s)=\dfrac{\partial L}{\partial\vec{r}'}-\dfrac{\mathrm{d}}{\mathrm{d}s}\left(\dfrac{\partial L}{\partial\vec{r}''}\right),&&\vec{m}(s)=\dfrac{\partial L}{\partial\vec{r}''},
\end{align}
where $\vec{r}(s)=(x(s),y(s))$. At the clamped end $s=0$, we impose $\vec{r}(0)=\vec{0},\vec{r}'(0)=\vec{\hat{x}}$, the unit vector in the $x$-direction. At the free end $s=1$ of the relaxed line, we impose $\vec{f}(1)=\vec{m}(1)=\vec{0}$. Let $\vec{R}(s)$ describe this relaxed line. For the compressed line, we then impose $\vec{r}(1)=\vec{R}(1-\delta)$ and $(\vec{r}'(1),z'(1))\parallel(\vec{R}'(1-\delta),z'(1-\delta))$. Because the latter are unit vectors by inextensibility, they are in fact equal, so $\vec{r}'(1)=\vec{R}'(1-\delta)$. With these conditions, $\mathcal{B}=0$.

Finally, the infinitesimal work done by the compressive forces $\vec{F_0}$ at $s\!=\!0$ and $\vec{F_1}$ at $s\!=\!1$ is ${\delta\mathcal{W}\!=\!\vec{F_0}\cdot\delta\vec{r}(0)\!+\!\vec{F_1}\cdot\delta\vec{r}(1)}$. By comparison with $\mathcal{B}$, we deduce $\vec{F_0}=\vec{f}(0)$, ${\vec{F_1}=-\vec{f}(1)}$.

\subsection{Asymptotic calculations} The full details of the asymptotic calculations are given in \hyperref[details]{Appendix B}.

\subsection{Numerical methods} We solve the Euler--Lagrange equation of \eqref{eq:L} numerically using the \texttt{bvp5c} solver of \textsc{Matlab} (The MathWorks, Inc.), relying on their first integral ${\lambda(s)=\lambda_0+3\kappa(s)^2/4}$, which is related to the tangential force balance on the elastic line~\cite{goldstein95}, and where $\lambda_0$ is a constant to be determined (\hyperref[details]{Appendix B}).

\section{\uppercase{Details of calculations}}\label{details}
This Appendix divides into four subsections: We first review Euler buckling on a flat surface. We then analyse ``extrinsic'' buckling of a short elastic line asymptotically, followed by a similar analysis of ``intrinsic'' buckling. Finally, we derive the solution of Lagrange multiplier function that we use in our numerical calculations.

\subsection{Flat Euler buckling}
We begin by reviewing the Euler buckling of a clamped inextensible elastic line of length $\ell$ in the plane (\textfigref{fig1}{A} of the main text). As in the main text, $s$ is arclength along the undeformed line, and its undeformed shape is $\bigl(x(s),y(s)\bigr)=(s,0)$. In what follows, we will scaling lengths by $\ell$, as announced in the main text. 

\onecolumngrid
\subsubsection*{\textbf{Buckling problem}} As introduced in the main text, we solve for the buckling of the elastic line upon compression by a relative amount $\delta$. The Lagrangian of the problem, corresponding to by~\eqref{eq:L} of the main text, becomes, after scaling lengths,
\begin{align}
    \mathcal{L} = \frac{1}{2}\int_0^1{\frac{x''(s)^2 + y''(s)^2}{\ell^2}\mathrm{d}s} - \int_0^1{\lambda(s)\bigl[x'(s)^2+y'(s)^2-1\bigr]\,\mathrm{d}s},
\end{align}
where dashes denote differentiation with respect to $s$, $x''(s)^2 + y''(s)^2$ is the squared curvature of the elastic line, and $\lambda(s)$ is the Lagrange multiplier function that enforces its inextensibility, $x'(s)^2+y'(s)^2=1$. Variation of the functional yields the Euler--Lagrange equations
\begin{subequations}\label{flateqs}
\begin{align}
\label{flateq1}
    \ell^{-2}x''''(s) + 2\lambda(s)x''(s) + 2\lambda'(s)x'(s) &= 0,\\
\label{flateq2}
    \ell^{-2}y''''(s) + 2\lambda(s)y''(s) + 2\lambda'(s)y'(s) &= 0,\\
\label{flateq3}
    x'(s)^2 + y'(s)^2 - 1 &= 0.
\end{align}
\end{subequations}
We write $\delta=\ell^2d$, with $d=O(1)$, and solve the buckling problem for $\ell\ll 1$. The flat buckling problem does not of course have any intrinsic length scale other than $\ell$, so the condition $\ell\ll 1$ only expresses the smallness of the relative compression $\delta=O\bigl(\ell^2\bigr)$ compared to $\ell$. The boundary conditions of clamped ends are, as discussed in the main text,
\begin{subequations}\label{flatbcs}
\begin{align}
\label{flatbc1}
    x(0)&=0,&y(0)&=0,&x'(0)&=1,&y'(0)&=0,\\
\label{flatbc2}
    x(1)&= 1-\ell^2d,&y(1)&= 0,&x'(1)&= 1,&y'(1) &= 0.
\end{align}
\end{subequations}
It is, in general, more convenient to express this flat Euler buckling problem in terms of the tangent angle $\theta(s)$, which satisfies $x'(s)=\cos{\theta(s)}$, $y'(s)=\sin{\theta(s)}$. However, the buckling problems on curved surfaces discussed in the next sections cannot be expressed easily in terms of this tangent angle, which is why we are solving the flat problem in terms of $x(s)$, $y(s)$, too.
\subsubsection*{\textbf{Asymptotic solution of the buckling problem}} To solve \eqsref{flateqs} subject to boundary conditions~\neqref{flatbcs} asymptotically, we seek a solution in the form
\begin{align}
    x(s) &= s + \ell^2 x_2(s) + O\bigl(\ell^4\bigr), &y(s) &= \ell y_1(s) + \ell^3y_3(s)+ O\bigl(\ell^5\bigr), &    \lambda(s) &= \ell^{-2}\lambda_{-2}+ \lambda_0(s) + O(\ell),
\end{align}
in which we have assumed that $\lambda_{-2}$ is constant, which follows immediately from \eqref{flateq1} at leading order.
As discussed in the main text, the boundary terms in the variation of the Lagrangian yield expressions for the forces at the ends of the elastic line. We thus find the forces at $s=0$ and $s=1$ to have components
\begin{subequations}\label{eq:fflat}
\begin{align}
    F_{0,x} &= -\frac{2\lambda_{-2}}{\ell^2}- \bigl[ 2\lambda_0(0)+x_2'''(0)\bigr]+O(\ell), &    F_{0,y} &= -\frac{y_1'''(0)}{\ell} -\ell y_3'''(0) + O\bigl(\ell^2\bigr),\\
    F_{1,x} &= -\frac{2\lambda_{-2}}{\ell^2}- \bigl[ 2\lambda_0(1)+x_2'''(1)\bigr]+O(\ell), &    F_{1,y} &= -\frac{y_1'''(1)}{\ell} -\ell y_3'''(1) + O\bigl(\ell^2\bigr).
\end{align}
\end{subequations}
\paragraph*{\textbf{Leading-order solution.}} At leading order, \eqref{flateq2} and boundary conditions~\neqref{flatbcs} give
\begin{align}
\lambda_\ast^2y_1''(s) + y_1''''(s) = 0\quad\text{subject to }y_1(0)=y_1'(0)=y_1(1)=y_1'(1)=0, \label{eq:eigflat}
\end{align}
where we have defined $\lambda_{-2} = \lambda_\ast^2/2$. The general solution of this fourth-order ordinary differential equation is
\begin{align}
    y_1(s) = a_1 + a_2 s + a_3 \sin{\lambda_\ast s} + a_4 \cos{\lambda_\ast s},
\end{align}
in which $a_1,a_2,a_3,a_4$ are constants that are determined by the boundary conditions via the system of linear equations
\begin{align}
\begin{pmatrix}
1 & 0& 0& 1 \\
1 & 1& \sin{\lambda_\ast}& \cos{\lambda_\ast} \\
0 & 1& \lambda_\ast& 0\\
0 & 1& \lambda_\ast\cos{\lambda_\ast}& -\lambda_\ast\sin{\lambda_\ast}
\end{pmatrix}\begin{pmatrix}
 a_1\\
 a_2\\
 a_3\\
 a_4
\end{pmatrix} = \begin{pmatrix}
 0\\
 0\\
 0\\
 0
\end{pmatrix}.
\end{align}
This has a nontrivial solution if and only if the determinant of the matrix on the left-hand side vanishes, which is if and only if\begin{subequations}
\begin{align}
    2\lambda_\ast\sin{\frac{\lambda_\ast}{2}}\left(\lambda_\ast\cos{\frac{\lambda_\ast}{2}}-2\sin{\frac{\lambda_\ast}{2}}\right)=0.
\end{align}
There are thus three cases,
\begin{align}
    &\text{(i)}\colon \lambda_\ast = 0,\qquad\text{(ii)}\colon \sin\frac{\lambda_\ast}{2} = 0,\qquad\text{and}\qquad\text{(iii)}\colon \tan{\frac{\lambda_\ast}{2}}=\frac{\lambda_\ast}{2}.\label{eq:lflat}
\end{align}\end{subequations}
Case (i) is actually the trivial solution in which the line remains flat, while cases (ii) and (iii) are symmetric and asymmetric buckled shapes, respectively. The corresponding shapes are
\begin{align}
\label{y1shapes}
    &y_1^{\text{(ii)}}(s) = a (1 - \cos{\lambda_\ast s}),&&
    y_1^{\text{(iii)}}(s) = a \left(1 - \cos{\lambda_\ast s}-2s+\frac{2 \sin{\lambda_\ast s}}{\lambda_\ast}\right),
\end{align}
in which $a=a_1$ is the buckling amplitude, to be determined. The first three of these modes are shown in \textfigref{fig1}{C} of the main text: $\lambda_\ast = 2\pi$ is the first buckled mode (orange line in \textfigref{fig1}{C}, symmetric); the second buckled mode is asymmetric, $\lambda_\ast\approx 2.86\pi$ yields the second and first asymmetric mode (green line in \textfigref{fig1}{C}); $\lambda_\ast=4\pi$ gives the third and second symmetric mode (blue line in \textfigref{fig1}{C}), and so on.

\paragraph*{\textbf{Symmetric solutions.}} We begin by analysing the symmetric solutions, for which, from the second of \eqsref{eq:lflat}, $\sin{\lambda_\ast}=0$ and $\cos{\lambda_\ast}=1$. We will use \text{Mathematica} (Wolfram, Inc.) throughout to handle the complicated algebraic expressions that arise in the subsequent calculations. With $\smash{y_1(s) = y_1^{\text{(ii)}}(s)}$, \eqref{flateq3} now gives 
\begin{align}
    2x_2'(s) + a^2\lambda_\ast^2\sin^2{\lambda_\ast s} = 0\quad\Longrightarrow\quad x_2(s) = -\frac{a^2\lambda_\ast}{8}(2\lambda_\ast s - \sin{2\lambda_\ast s}),\label{eq:flatx2}
\end{align}
using $x_2(0)=0$ from the first of boundary conditions \neqref{flatbc1}. Now the first of \eqsref{flatbc2} implies $x_2(1) =-d$, which determines two solutions for the buckling amplitude,
\begin{align}
\label{Amplitude}
    a = \pm\frac{2\sqrt{d}}{\lambda_\ast},
\end{align}
corresponding to ``up'' and ``down'' buckling, respectively, which are equivalent for a flat surface. Meanwhile, integrating \eqref{flateq1}, we find
\begin{align}
    \lambda_0(s) = C_{\lambda_0} + \frac{3a^2\lambda_\ast^4}{8}\cos{2\lambda_\ast s},
\end{align}
where $C_{\lambda_0}$ is a constant of integration that remains to be determined. To do so, we return to \eqref{flateq2}, which gives
\begin{subequations}
\begin{align}
    \frac{a\lambda_\ast^2}{4}\cos{\lambda_\ast s}\bigl( 8C_{\lambda_0} - 6a\lambda_\ast^4 + 9a^2\lambda^4\cos{2\lambda_\ast s}\bigr) + \lambda_\ast^2y_3''(s) + y_3''''(s) = 0.
\end{align}
Boundary conditions~\neqref{flatbcs} imply $y_3(0) = y_3(1) = y_3'(0) = y_3'(1) = 0$, whence
\begin{align}
    &y_3(s)= -\dfrac{a^3\lambda_\ast^2}{16}(2+\cos{\lambda_\ast s})(1-\cos{\lambda_\ast s})^2+C_{y_3}y_1^{\text{(ii)}}(s),&&C_{\lambda_0} = \frac{3a^2\lambda_\ast^4}{16},
\end{align}
\end{subequations}
where $C_{y_3}$ is another constant of integration which we will not determine, because we can now compute, from \eqsref{eq:fflat},
\begin{subequations}
\begin{align}
    F_{0,x} &= -\frac{\lambda_\ast^2}{\ell^2} - \frac{a^2\lambda_\ast^4}{8} + O(\ell) = -\frac{\lambda_\ast^2}{\ell^2} - \frac{d\lambda_\ast^2}{2} + O(\ell)  = -F_{1,x},&&F_{0,y} = O\bigl(\ell^2\bigr) = -F_{1,y}.
\end{align}
The total buckling force is therefore
\begin{align}
    F_1 = \frac{\lambda_\ast^2}{\ell^2} + \dfrac{d\lambda_\ast^2}{2} + O(\ell) = F_0.\label{eq:fflatsym}
\end{align}
\end{subequations}
\paragraph*{\textbf{Asymmetric solutions.}} Next, we analyse the asymmetric buckling solutions. The last of \eqsref{eq:lflat} implies
\begin{align}
\sin{\lambda_\ast}&=\dfrac{4\lambda_\ast}{4+\lambda_\ast^2},&\cos{\lambda_\ast}&=\dfrac{4-\lambda_\ast^2}{4+\lambda_\ast^2},\label{eq:asymsimp}
\end{align}
which will allow us to simplify the results of our calculations. Taking $\smash{y_1(s)=y_1^{\text{(iii)}}(s)}$, \eqref{flateq3} gives
\begin{subequations}
\begin{align}
    a^2(\lambda_\ast \sin{\lambda_\ast s}+2 \cos{\lambda_\ast s}-2)^2+2 x_2'(s) = 0,
\end{align}
from which we obtain, using $x_2(0)=0$ as in the symmetric case,
\begin{align}
    x_2(s) = \frac{a^2}{8\lambda_\ast}\left(12\lambda_\ast - 2 \lambda_\ast^3 s + \lambda_\ast^2\sin{2\lambda_\ast s} - 24\lambda_\ast s + 32\sin{\lambda_\ast s} - 4\sin{2 
    \lambda_\ast s} - 16\lambda_\ast\cos{\lambda_\ast s} + 4\lambda_\ast\cos{2\lambda_\ast s}\right).
\end{align}
\end{subequations}
Again, as in the symmetric case, $x_2(1)=-d$. It follows that \eqref{Amplitude} continues to hold. Similarly, integrating \eqref{flateq1}, we find
\begin{align}
    \lambda_0(s) = C_{\lambda_0}+\frac{3a^2\lambda_\ast^2}{8} \left(\lambda_\ast^2 - 4\right)\cos{2\lambda_\ast s} - \frac{3a^2\lambda_\ast^3}{2}\sin{2\lambda_\ast s}.
\end{align}
To determine the constant of integration $C_{\lambda_0}$, as in the symmetric case, we need to determine $y_3(s)$. From \eqref{flateq3}, we have
\begin{subequations}
\begin{align}
    &\frac{a\lambda_\ast}{4}(\lambda_\ast\cos{\lambda_\ast s} - 2\sin{\lambda_\ast s})\left\{8C_{\lambda_0}+3a^2\lambda_\ast^2\left[3\bigl(\lambda_\ast^2 - 4\bigr) \cos{2\lambda_\ast s}+8\lambda_\ast\sin{\lambda_\ast s}- 12\lambda_\ast\sin{2\lambda_\ast s} + 16 \cos{\lambda_\ast s}-2\bigl(\lambda_\ast^2 + 4\bigr)\right]\right\} \nonumber\\
    &\qquad\qquad + \lambda_\ast^2 y_3''(s) + y_3''''(s) = 0,
\end{align}
subject to $y_3(0) = y_3(1) = y_3'(0) = y_3'(1) = 0$ as in the symmetric case, which yields
\begin{align}
    y_3(s)&= \dfrac{a^3}{64\lambda_\ast}\left[(1-2s)\lambda_\ast\bigl(\lambda_\ast^2+132\bigr)-16 \bigl(\lambda_\ast^2-2 \lambda_\ast^2 s+14\bigr)\sin{\lambda_\ast s}-16 \bigl(\lambda_\ast^2-4\bigr) \sin{2 \lambda_\ast s}+2 \bigl(3 \lambda_\ast^2-4\bigr) \sin{3 \lambda_\ast s}\right.\nonumber\\
    &\qquad\qquad-\left.\lambda_\ast \bigl(\lambda_\ast^2-12\bigr) \cos{3 \lambda_\ast s}-16 \lambda_\ast (4 s+5) \cos{\lambda_\ast s}-64 \lambda_\ast \cos{2 \lambda_\ast s}\right]+C_{y_3}y_1^{\text{(iii)}}(s),\\
    C_{\lambda_0} &= \frac{a^2\lambda_\ast^2\bigl(4 + 3\lambda_\ast^2\bigr)}{16},
\end{align}
\end{subequations}
where $C_{y_3}$ is, again, a constant of integration. The forces at the ends of the rod are now, from \eqref{eq:fflat},
\begin{subequations}
\begin{align}
    F_{0,x} &= -\frac{\lambda_\ast^2}{\ell^2} - \frac{a^2\lambda_\ast^2(\lambda_\ast^2 - 20)}{8} + O(\ell) = -F_{1,x},&    F_{0,y} &= \frac{2a\lambda_\ast^2}{\ell} + O(\ell) = -F_{1,y}.
\end{align}
The total buckling force is therefore

\vspace{-20pt}
\begin{align}
    F_1 = \frac{\lambda_\ast^2}{\ell^2} + \frac{d(\lambda_\ast^2 - 4)}{2} + O(\ell) = F_0.\label{eq:fflatasym}
\end{align}
\end{subequations}
Equations~\neqref{eq:fflatsym} and \neqref{eq:fflatasym} define the bifurcation diagram for flat Euler buckling, as shown in \textfigref{fig1}{C} of the main text. One can of course continue this expansion easily to any asymptotic order, but this will not be needed for the purposes of this paper.

\subsection{(Extrinsic) Euler buckling on curved surfaces}
As in the main text, we consider an inextensible elastic line of length $\ell\ll 1$ confined to a smooth curved surface $z=h(x,y)$ in Cartesian coordinate axes and clamped at one end (\textfigref{fig1}{B} of the main text), and its buckling under compression and clamping of the other end along its initial relaxed shape. Up to translation and rotation of the surface, the initial clamping is at the origin of the coordinate axes, in the $x$-direction, and the surface is tangent to the $(x,y)$ plane there. Scaling lengths by $\ell$ again, the surface therefore has a generic expansion
\begin{align}
h(x,y)&=\ell\bigl(b_{20}x^2\!+\!b_{11} xy \!+\! b_{02} y^2\bigr)\!+\!\ell^2\bigl(c_{30} x^3 \!+\! c_{21} x^2 y\!+\! c_{12}x y^2 \!+\! c_{03} y^3\bigr)\!+\!\ell^3\bigl(d_{40} x^4 \!+\! d_{31}x^3y \!+\! d_{22}x^2y^2 \!+\! d_{13}xy^3 \!+\! d_{04}y^4\bigr)\!+\!O\bigl(\ell^4\bigr), \label{eq:h}
\end{align}
in which we assume that $b_{20},b_{11},b_{02},c_{30},c_{21},c_{12},c_{03},d_{40},d_{31},d_{22},d_{13},d_{04}=O(1)$. We stress that the condition $\ell\ll1$ is no longer simply a condition imposing the smallness of the relative compression, as in the flat case, but now expresses the shortness of the elastic line compared to the additional length scales in the problem set by the curvature of the surface. To describe the buckling of the elastic line on this surface, we need to determine, first, its relaxed shape (which is not, in general, a geodesic of the surface), and then solve the buckling problem. A point on the elastic line has coordinates $\bigl(x(s),y(s),z(s)\bigr)$ where $s\in[0,1]$ is arclength and $\smash{z(s)=h\bigl(x(s),y(s)\bigr)}$. The Lagrangian of the problem is, after scaling lengths by $\ell$,
\begin{align}
\label{L_complete}
    &\mathcal{L}=\int_0^1{L\bigl(x(s),y(s),x'(s),y'(s),x''(s),y''(s),\lambda(s)\bigr)\,\mathrm{d}s}\nonumber\\
    &\qquad\text{with } L\bigl(x(s),y(s),x'(s),y'(s),x''(s),y''(s),\lambda(s)\bigr)=\frac{x''(s)^2+y''(s)^2+z''(s)^2}{2\ell^2} - \lambda(s)\bigl[x'(s)^2+y'(s)^2+z'(s)^2-1\bigr],
\end{align}
where dashes again denote differentiation with respect to arclength, in which, as in \eqref{eq:L} of the main text, $\lambda(s)$ is the Lagrange multiplier function that imposes inextensibility, $x'(s)^2+y'(s)^2+z'(s)^2=1$. We note that an explicit dependence of the Lagrangian density $L$ on $x(s),y(s)$ arises via $z(s)=h\bigl(x(s),y(s)\bigr)$.
\subsubsection*{\textbf{Euler--Lagrange equations}}
We begin by deriving the Euler--Lagrange equations that describe both the relaxed line and its buckling. The variation of the Lagrangian~\neqref{L_complete} is
\begin{align}
    \delta \mathcal{L} &= \int_0^1{\bigl(L_{,x} \delta x + L_{,x'}\delta x' + L_{,x''}\delta x'' + L_{,y}\delta y + L_{,y'}\delta y' + L_{,y''}\delta y'' + L_{,\lambda}\bigr)\,\mathrm{d}s}\nonumber\\
    &=\left\llbracket L_{,x''}\delta x'+L_{,y''}\delta y'+\biggl(L_{,x'} - \frac{\mathrm{d}L_{,x''}}{\mathrm{d}s}\biggr)\delta x+\left(L_{,y'} - \frac{\mathrm{d}L_{,y''}}{\mathrm{d}s}\right)\delta y\right\rrbracket_0^1\nonumber\\
    &\qquad+\int_0^1{\left[\left(L_{,x} - \frac{\mathrm{d}L_{,x'}}{\mathrm{d}s} + \frac{\mathrm{d}^2L_{,x''}}{\mathrm{d}s^2}\right)\delta x+\left(L_{,y} - \frac{\mathrm{d}L_{,y'}}{\mathrm{d}s} + \frac{\mathrm{d}^2L_{,y''}}{\mathrm{d}s^2}\right)\delta y+L_{,\lambda}\delta\lambda\right]\mathrm{d}s},
\end{align}
where
\begin{align}
&L_{,x}=\dfrac{\partial L}{\partial x},&&L_{,x'}=\dfrac{\partial L}{\partial x'},&&L_{,x''}=\dfrac{\partial L}{\partial x''},&L_{,y}=\dfrac{\partial L}{\partial y},&&L_{,y'}=\dfrac{\partial L}{\partial y'},&&L_{,y''}=\dfrac{\partial L}{\partial y''},&&L_{,\lambda}=\dfrac{\partial L}{\partial\lambda}.
\end{align}
We read the Euler--Lagrange equations off this variation, viz.,
\begin{subequations}\label{EL_complete}
\begin{align}
\label{EL_complete x}
    L_{,x} - \frac{\mathrm{d}L_{,x'}}{\mathrm{d}s} + \frac{\mathrm{d}^2L_{,x''}}{\mathrm{d}s^2} = 0, \\
\label{EL_complete y}
    L_{,y} - \frac{\mathrm{d}L_{,y'}}{\mathrm{d}s} + \frac{\mathrm{d}^2L_{,y''}}{\mathrm{d}s^2} = 0, \\
\label{EL_complete lambda}
    L_{,\lambda} = 0,
\end{align}
\end{subequations}
\subsubsection*{\textbf{Asymptotic calculation of the shape of the relaxed elastic line}} 
We can now solve the Euler--Lagrange equations~\neqref{EL_complete} for a relaxed elastic line by asymptotic expansion in $\ell\ll 1$. The boundary conditions are
\begin{align}
&x=0,\;y=0,\;x'=1,\;y'=0\quad\text{at }s=0,&&L_{,x'} - \frac{\mathrm{d}L_{,x''}}{\mathrm{d}s}=0,\;L_{,y'} - \frac{\mathrm{d}L_{,y''}}{\mathrm{d}s}=0,\;L_{,x''}=0,\;L_{,y''}=0\quad\text{at }s=1,\label{BC_relaxed}
\end{align}
which can be read off the variation as discussed in the main text. Using \textsc{Mathematica} to expand the Euler--Lagrange equations~\neqref{EL_complete} and boundary conditions~\neqref{BC_relaxed}, we seek a solution in the form
\begin{subequations}
\begin{align}
    x(s) &= s + \ell x_1(s) + \ell^2 x_2(s) + \ell^3 x_3(s) + \ell^4 x_4(s) + O\bigl(\ell^5\bigr), \\
    y(s) &= \ell y_1(s) + \ell^2 y_2(s) + \ell^3 y_3(s) + \ell^4 y_4(s) + O\bigl(\ell^5\bigr), \\
    \lambda(s) &= \ell^{-2}\lambda_{-2}(s) + \ell^{-1}\lambda_{-1}(s) + \lambda_0(s) + \ell\lambda_{1}(s) + \ell^{2}\lambda_{2}(s) + O\bigl(\ell^3\bigr).
\end{align}
\end{subequations}
\paragraph*{\textbf{Solution at order $\vec{O(\ell)}$.}} At leading order, the Euler--Lagrange equations yield
\begin{align}
&x_1'(s)= 0,&&\lambda_{-2}'(s)=\lambda_{-1}'(s)=0,&&y_1''''(s)=0,
\end{align}
while boundary conditions give
\begin{align}
&x_1(0)=0,&&\lambda_{-2}(1)=\lambda_{-1}(1)=0,&&y_1(0)=y_1'(0)=y_1''(1)=y_1'''(1)=0.
\end{align}
The solution of these differential equations is
\begin{align}
&x_1(s)=0,&&\lambda_{-2}(s)=\lambda_{-1}(s)=0,&&y_1(s)=0.
\end{align}
\paragraph*{\textbf{Solution at order $\vec{O(\ell^2)}$.}} At next order, we now find
\begin{align}
&4b_{20}^2s^2+2x_2'(s) = 0,&&\lambda_0'(s)=0,&&y_2''''(s) = 0,
\end{align}
subject to
\begin{align}
&x_2(0) = 0,&&\lambda_0(1) = 4b_{20}^2,&&y_2(0) = y_2'(0)=0,\quad y_2''(1) = -2b_{11}b_{20},\quad y_2'''(1) = 2b_{11}b_{20}.
\end{align}
The solution is
\begin{align}
&x_2(s) = -\frac{2b_{20}^2}{3}s^3,&&\lambda_0(s) = 4b_{20}^2,&&y_2(s) = \frac{b_{11}b_{20}}{3}(s-6)s^2.
\end{align} 
\paragraph*{\textbf{Solution at order $\vec{O(\ell^3)}$.}} Continuing the expansion, we obtain the differential equations
\begin{align}
    &12b_{20}c_{30}s^3 + 2x_3'(s) = 0,&&-36b_{20}c_{30} + 2\lambda_1'(s) = 0,&&y_3''''(s) = 0,
\end{align}
subject to
\begin{align}
&x_3(0) = 0,&&\lambda_1(1) = 24b_{20}c_{30},&&y_3(0)=y_3'(0)=0,\quad y_3''(1)=-2b_{20}c_{21}-6b_{11}c_{30},\quad y_3'''(1)=4b_{20}c_{21}.
\end{align}
This gives
\begin{align}
&x_3(s) = -\frac{3b_{20}c_{30}}{2}s^4,&&\lambda_1(s)=6b_{20}c_{30}(3s + 1),&&y_3(s)=\dfrac{s^2}{3}\bigl[b_{20}c_{21}(2s-9)-9b_{11}c_{30}\bigr].
\end{align}
\paragraph*{\textbf{Solution at order $\vec{O(\ell^4)}$.}} Finally, we obtain differential equations for $x_4(s)$, $\lambda_2(s)$, and $y_4(s)$,
\begin{subequations}
\begin{align}
\frac{s^2}{3}\left[\bigl(-52b_{20}^4+27c_{30}^2+48b_{20}d_{40}\bigr)s^2+b_{11}^2b_{20}^2\bigl(48-96s+19s^2\bigr)\right] + 2x_4'(s) &= 0,\\
-6b_{11}^2b_{20}^2(5s - 8) - 18s\bigl(-4b_{20}^4 + 4b_{20}d_{40} + 3c_{30}^2\bigr) + \lambda_2'(s) &= 0,\\
8b_{11}\bigl(-4b_{20}^3+b_{11}^2b_{20}s+3d_{40}s\bigr)+y_4''''(s) &= 0.
\end{align}
\end{subequations}
The boundary conditions are $x_4(0)=y_4(0)=y_4'(0)=0$ and
\begin{subequations}
\begin{align}
    \lambda_2(1) &=- 28b_{11}^2b_{20}^2 - 48b_{20}^4 + 48 b_{20}d_{40} + 36c_{30}^2,\\
    y_4''(1)&=8b_{11}^3b_{20}-6c_{21}c_{30}-2b_{20}d_{31}+\frac{4b_{11}}{3}\bigl(5b_{02}b_{20}^2+13b_{20}^3-9d_{40}\bigr),\\
    y_4'''(1)&=-4b_{11}^3b_{20} + 6c_{21}c_{30} +6b_{20}d_{31} - 4b_{11}\bigl(3b_{02}b_{20}^2-5b_{20}^3+3d_{40}\bigr).
\end{align}
\end{subequations}
It follows that
\begin{subequations}
\begin{align}
    x_4(s) &= -\frac{s^3}{30}\left[b_{11}^2b_{20}^2\bigl(80-120s+19s^2\bigr)+\bigl(-52b_{20}^4+27c_{30}^2+48b_{20}d_{40}\bigr)s^2\right],\\ 
    y_4(s) &=-\frac{s^2}{15}\left\{b_{11}^3b_{20}\bigl(s^3-70\bigr)+ b_{11}\left[10b_{02}b_{20}^2(3s-14)-10b_{20}^3\bigl(2s^2-3s+10\bigr) +3d_{40}\bigl(s^3+20\bigr)\right]\right.\nonumber\\
    &\qquad\qquad-\left.15\bigl[c_{21}c_{30}(s-6)+b_{20}d_{31}(s-4)\bigr] \right\},\\
    \lambda_2(s) &= b_{11}^2b_{20}^2\bigl(15s^2 - 48s + 5\bigr) - 3\bigl(4 b_{20}^4 - 4 b_{20}d_{40} - 3 c_{30}^2\bigr)\bigl(3s^2 + 1\bigr). 
\end{align}
\end{subequations}
It is easy to check that these solutions satisfy the remaining boundary conditions that we have not explicitly imposed in the above calculations. The shape and tension of a (short) relaxed elastic line are thus described by
\begin{subequations}\label{relaxed elastic line}
\begin{align}
\label{relaxed elastic line x}
    X(s) &= s - \frac{2b_{20}^2s^3}{3}\ell^2 -\frac{3b_{20}c_{30}s^4}{2}\ell^3 -\frac{s^3}{30}\left[b_{11}^2b_{20}^2\bigl(80-120s+19s^2\bigr)+\bigl(-52b_{20}^4+27c_{30}^2+48b_{20}d_{40}\bigr)s^2\right]\ell^4 + O\bigl(\ell^5\bigr), \\
\label{relaxed elastic line y}
    Y(s) &= \frac{s^2}{3}b_{11}b_{20}(s-6)\ell^2 +\dfrac{s^2}{3}\bigl[b_{20}c_{21}(2s-9)-9b_{11}c_{30}\bigr]\ell^3-\frac{s^2}{15}\left\{b_{11}^3b_{20}\bigl(s^3-70\bigr)+ b_{11}\left[10b_{02}b_{20}^2(3s-14)\right.\right.\nonumber\\
    &\quad\quad\left.\left.-10b_{20}^3\bigl(2s^2-3s+10\bigr) +3d_{40}\bigl(s^3+20\bigr)\right]-15\bigl[c_{21}c_{30}(s-6)+b_{20}d_{31}(s-4)\bigr] \right\}\ell^4+ O\bigl(\ell^5\bigr), \\
\label{relaxed elastic line lambda}
    \Lambda(s) &= 4b_{20}^2 + 6b_{20}c_{30}(3s + 1)\ell + \left[b_{11}^2b_{20}^2\bigl(15s^2 - 48s + 5\bigr) - 3\bigl(4 b_{20}^4 - 4 b_{20}d_{40} - 3 c_{30}^2\bigr)\bigl(3s^2 + 1\bigr)\right]\ell^2 + O(\ell^3).
\end{align}
\end{subequations}

\subsubsection*{\textbf{Asymptotic buckling of a short relaxed elastic line}} As described in the main text, we now clamp the free end of the relaxed elastic line along its relaxed shape and compress it by a relative amount $\delta$. We solve this buckling problem by solving the Euler--Lagrange equations~\neqref{EL_complete}, replacing the boundary conditions at $s=1$ in \eqsref{BC_relaxed} with
\begin{align}
    x(1) = X(1-\delta),\quad x'(1) = X(1-\delta),\quad y(1) = Y(1-\delta),\quad y'(1) = Y'(1-\delta),\label{BC_buckling}
\end{align}
as discussed in the Materials \& Methods of the main text. We first solve the buckling problem for $\delta=\ell^2d$, with $d=O(1)$, as in the case of flat buckling. We make the buckling ansatz
\begin{subequations}\label{eq:ansatz}
\begin{align}
    x(s) &= X(s) + \ell^2 x_{2}(s) + \ell^3x_{3}(s) + O\bigl(\ell^4\bigr),\\
    y(s) &= Y(s) + \ell y_{1}(s) + \ell^2y_{2}(s) + \ell^3y_{3}(s) + O\bigl(\ell^4\bigr),\\
    \lambda(s) &=  \Lambda(s) + \ell^{-2}\lambda_{-2}(s) + \ell^{-1}\lambda_{-1}(s) + \lambda_0(s) + O(\ell). 
\end{align}
\end{subequations}
\paragraph*{\textbf{Leading-order solution.}} As in the flat case, we solve the buckling problem order-by-order. From \eqref{EL_complete x}, $\lambda_{-2},\lambda_{-1}$ are constants. On setting $\lambda_{-2}=\lambda_\ast^2/2$, \eqref{EL_complete y} yields
\begin{align}
    \lambda_\ast^2y_1''(s) + y_1''''(s) = 0,\quad\text{subject to }y_1(0)=y_1'(0)=y_1(1)=y_1'(1)=0. 
\end{align}
This is the same eigenvalue problem that we found above, as \eqref{eq:eigflat}, for flat buckling. Its eigenvalues are given by \eqref{eq:lflat}, with corresponding symmetric and asymmetric buckling modes given by \eqsref{y1shapes}. 

\paragraph*{\textbf{Symmetric buckling.}} Again, we begin by taking $y_1(s)=y_1^{\text{(ii)}}(s)$ and analysing the symmetric solutions. \eqref{EL_complete lambda} gives, identically to \eqref{eq:flatx2},
\begin{align}
2x_2'(s)+a^2\lambda_\ast^2\sin^2{\lambda_\ast s} = 0\quad\Longrightarrow\quad x_2(s) = -\frac{a^2 \lambda_\ast }{8}(2\lambda_\ast s-\sin{2\lambda_\ast s}),
\end{align}
using $x_2(0)=0$. This solution also satisfies $x_2'(0)=x_2'(1)=0$, while the remaining boundary condition, $x_2(1)=-d$, yields
\begin{align}
a=a_\pm\equiv\pm\dfrac{2\sqrt{d}}{\lambda_\ast},\label{eq:a}
\end{align}
again as in the flat case. Equation~\neqref{EL_complete x} now gives 
\begin{align}
    \frac{3a^2\lambda_\ast^5}{2}\sin{2\lambda_\ast s} + 2\lambda_0'(s) =0\quad \Longrightarrow \quad \lambda_0(s) = C_{\lambda_0} + \frac{3a^2\lambda_\ast^4}{8}\cos{2\lambda_\ast s},
\end{align}
where, again, the constant of integration $C_{\lambda_0}$ is undetermined at this order. The curvature of the surface finally comes, and the solution begins to differ from the flat buckling solution, at next order, where \eqref{EL_complete y} gives
\begin{align}
    2\lambda_\ast^2\bigl[2b_{11}b_{20}(s-1) + a\lambda_{-1}\cos{\lambda_\ast s}\bigr] + \lambda_\ast^2y_2''(s) + y_2''''(s)=0,
\end{align}
subject to $y_2(0)=y_2'(0)=y_2(1)=y_2'(1)=0$, whence
\begin{align}
&y_2(s) = \frac{b_{11}b_{20}}{3\lambda_\ast^2}\left\{2\left[9-\lambda_\ast^2(s - 2)(s - 1)s\right] + 2\lambda_\ast(2 - 3s)\sin{\lambda_\ast s}-18\cos{\lambda_\ast s}\right\}+C_{y_2}y_1^{\text{(ii)}}(s),&&\lambda_{-1} = -\frac{2b_{11}b_{20}}{a},
\end{align}
where $C_{y_2}$ is another constant of integration. Thus $\lambda_{-1}$ blows up as $a\to 0$ or, equivalently, $d\to 0$, unless $b_{20}b_{11}=0$. This is precisely the condition noted in the main text: The curved buckling problem is uninteresting unless $b_{20}b_{11}\neq 0$. We can now continue the expansion to order $O\bigl(\ell^3\bigr)$. From \eqref{EL_complete y}, we obtain
\begin{align}
&\lambda_\ast^2 \left\{4ab_{02}b_{20} + 2(s - 1)\left[3b_{11}c_{30}(s + 1)+b_{20}c_{21} (s+3)\right]\right\}-\frac{\cos{\lambda_\ast s}}{8a}\left\{3a^4 \lambda_\ast^6+16a\lambda_\ast^2 \left[2ab_{20} (b_{02}-2b_{20})-aC_{\lambda_0}+2b_{11}b_{20} C_{y_2}\right]\right.\nonumber\\
&\quad+\left.64b_{11}^2b_{20}^2\right\}+\frac{9a^3\lambda_\ast^6}{8} \cos{3\lambda_\ast s}-\frac{2 b_{11}^2\lambda_\ast\sin{\lambda_\ast s}}{3 a}\left[3a^2 \lambda_\ast^2 s+4b_{20}^2 (3s-2)\right]+\lambda_\ast^2y_3''(s)+y_3''''(s)=0,
\end{align}
subject to $y_3(0)=y_3'(0)=y_3(1)=y_3'(1)=0$, with solution
\begin{subequations}
\begin{align}
y_3(s)& = \frac{1}{192a\lambda_\ast^4}\left[3a^4 \lambda_\ast^6(1-\cos{3\lambda_\ast s}) + 16 \cos{\lambda_\ast s} \Bigl(b_{11}^2 \left\{3 a^2 \lambda_\ast^2\bigl(2 \lambda_\ast^2 s^2+1\bigr)+4 b_{20}^2 \left[2 \lambda_\ast^2 s (3 s-4)+3\right]\right\}\right.\nonumber\\ 
    &\hspace{18mm}+24 a b_{20} \left[\lambda_\ast^2 (6 a b_{02}-5 c_{21})-6 c_{21}\right]-144 a b_{11} c_{30}\bigl(\lambda_\ast^2+3\bigr)\Bigr)-16 b_{11}^2 \left[3 a^2 \lambda_\ast^2\bigl(2\lambda_\ast^2 s+1\bigr)+4b_{20}^2\bigl(3-2\lambda_\ast^2 s\bigr)\right]\nonumber\\
    &\hspace{18mm}-32 \lambda_\ast \sin{\lambda_\ast s} \left\{a\lambda_\ast^2 \left[3a(2s - 1)\bigl(b_{11}^2 - 4b_{02}b_{20}\bigr) + 3b_{11}c_{30} (8s - 5) + b_{20}c_{21}(20s - 13)\right]\right.\nonumber\\
    &\hspace{18mm}+\left.12(2s - 1)\bigl(3ab_{11} c_{30}
    +a b_{20}c_{21}+b_{11}^2 b_{20}^2\bigr)\right\}+96 a b_{11} c_{30} \left[72+12\lambda_\ast^2\bigl(s^2-s+2\bigr)-\lambda_\ast^4 s\bigl(s^3-6 s+5\bigr)\right]\nonumber\\
    &\hspace{18mm}+32 a b_{20} \left\{72c_{21}+ 12\lambda_\ast^2\left[c_{21}\bigl(s^2-s + 5\bigr) - 6a b_{02}\right]-\lambda_\ast^4s(s - 1)\left[12ab_{02} + c_{21}\bigl(s^2+5s - 13\bigr)\right]\right\}\Bigr]\nonumber\\
    &\qquad+ C_{y_3}y_1^{\text{(ii)}}(s),\\
    C_{\lambda_0} &= \dfrac{1}{48a^2}\biggl\{8a \left[12ab_{20} (3b_{02} - 2b_{20})+9ab_{11}^2+4 (3 b_{11}b_{20}C_{y_2}- 6b_{11}c_{30}-5b_{20}c_{21})\right]+9a^4\lambda_\ast^4\nonumber\\
    &\qquad\qquad-\left.\dfrac{96}{\lambda_\ast^2}\bigl(6ab_{11}c_{30}+2ab_{20}c_{21}-5b_{11}^2 b_{20}^2\bigr)\right\},
\end{align}
\end{subequations}
in which $C_{y_3}$ is yet another constant of integration. Unlike in the flat buckling problem, we are not done at this stage yet, however, because it turns out that we need to determine $C_{y_2}$. We do so by expanding \eqref{EL_complete y}, which yields
\begin{align}
    &a b_{11}b_{20}\left\{\sin{\lambda_\ast s}\left[12\sin{\lambda_\ast s} + \lambda_\ast\bigl(3s^2 - 4\bigr)\right]+ \lambda_\ast (2 - 3s)\sin{2\lambda_\ast s} + 6s(1-\cos{\lambda_\ast s})\right\}+ a^2C_{y_2}\lambda_\ast^2\sin{\lambda_\ast s}+3x_3'(s) = 0.
\end{align}
Imposing $x_3(0)=x_3(1) = 0$ gives
\begin{align}
    &x_3(s) = \frac{2ab_{11}b_{20}}{3}\sin^2\left(\dfrac{\lambda_\ast s}{2}\right)\left[(3s - 2)\cos{\lambda_\ast s} - 3s^2+3s+ 2\right],&&C_{y_2}=-\dfrac{5b_{11}b_{20}}{a\lambda_\ast^2}.
\end{align}
We can now, finally, determine the buckling forces $F_0$ and $F_1$ at $s=0$ and $s=1$, respectively. We find
\begin{subequations}
\begin{align}
F_0&=\frac{\lambda_\ast^2}{\ell ^2} - \dfrac{1}{\ell}\left(\frac{4b_{11}b_{20}}{a}\right) + \left[\frac{a^2 \lambda_\ast^4}{8}+3b_{11}^2+4b_{20}(2b_{20}-3b_{02})-\dfrac{4b_{20}c_{21}\left(5\lambda_\ast^2+24\right)+24b_{11}c_{30}\left(\lambda_\ast^2+3\right)}{3a\lambda_\ast^2}\right] + O(\ell),\\
F_1&=F_0+2b_{20}^2\lambda_\ast^2+O(\ell),
\end{align}
\end{subequations}
Hence, if $b_{11}b_{20}=0$, ``up'' and ``down'' buckling are equivalent (to this order of expansion). If $b_{11}b_{20}\neq 0$, the lowest buckling force is selected. Thus $a=a_\pm$ if $b_{11}b_{20}\gtrless 0$. In particular, this shows that
\begin{align}
F_1&=\frac{\lambda_\ast^2}{\ell^2} -\dfrac{1}{\ell}\left(\frac{2\left|b_{11}b_{20}\right|\lambda_\ast}{\sqrt{d}}\right) + \left\{ \frac{d\lambda_\ast^2}{2}+3b_{11}^2+2b_{20}^2\lambda_\ast^2+4b_{20}(2b_{20}-3b_{02})\right.\nonumber\\
&\hspace{45mm}-\left.\operatorname{sign}(b_{11}b_{20})\left[ \frac{4(3b_{11}c_{30} + b_{20}c_{21})}{\sqrt{d} \lambda_\ast}+\frac{2\lambda_\ast (6b_{11}c_{30} + 5b_{20}c_{21})}{3\sqrt{d}}\right]\right\} + O(\ell).\label{eq:F1sym}
\end{align}
For the lowest buckling mode with $\lambda_\ast=2\pi$, for $c_{30}=c_{21}=0$, and identifying $b_{20}=A,b_{11}=B,b_{02}=C,d=d_{(2)}$, this establishes \eqref{eq:F2} of the main text. Equation~(\ref{eq:F1sym}) stresses however, that, for a general surface, corrections to the quadratic surface studied in the main text already enter at this order of expansion.

\paragraph*{\textbf{Asymmetric buckling.}} Next, we take $y_1(s) = y_1^{\text{(iii)}}(s)$ to analyse asymmetric buckling. We will use \eqsref{eq:asymsimp} throughout to simplify the results. Identically to the flat case, \eqref{EL_complete lambda} gives
\begin{align}
    a^2(\lambda_\ast\sin{\lambda_\ast s} + 2\cos{\lambda_\ast s} - 2)^2 +2x_2'(s) = 0,
\end{align}
so, with $x_2(0)=0$, we find, again,
\begin{align}
    x_2(s) = \frac{a^2}{8 \lambda_\ast}\left\{\bigl(\lambda_\ast^2-4\bigr) \sin{2\lambda_\ast s} + 32\sin{\lambda_\ast s} - 16\lambda_\ast\cos{\lambda_\ast s} + 4\lambda_\ast\cos{2\lambda_\ast s}-2\lambda_\ast \left[\bigl(\lambda_\ast^2+12\bigr) s-6\right]\right\}.
\end{align}
Again as in the flat case, requiring $x_2(1)=-d$ gives the relation between $d$ and the amplitude $a$ and $d$, which is again given by \eqref{eq:a}. Next, \eqref{EL_complete x} yields
\begin{align}
    \frac{3a^2\lambda_\ast^3}{2} \left[\bigl(\lambda_\ast^2-4\bigr)\sin{2\lambda_\ast s}+4\lambda_\ast \cos {2\lambda_\ast s}\right]+ 2\lambda_0'(s) = 0,
\end{align}
so, identically to the flat case,
\begin{align}
    \lambda_0(s) = C_{\lambda_0} + \frac{3}{8}a^2\lambda_\ast^2 \left[\bigl(\lambda_\ast^2 - 4\bigr)\cos{2\lambda_\ast s} - 4\lambda_\ast\sin{2\lambda_\ast s}\right],
\end{align}
where $C_{\lambda_0}$ is a constant of integration to be determined. As in the symmetric case, the curved surface geometry first enters the problem in the equation for $y_2$, which we obtain from \eqref{EL_complete y},
\begin{align}
2\lambda_\ast\left[-2a\lambda_{-1}\sin{\lambda_\ast s} + a\lambda_\ast\lambda_{-1}\cos{\lambda_\ast s} + 2b_{11}b_{20}\lambda_\ast(s - 1)\right] + \lambda_\ast^2y_2''(s) + y_2''''(s)= 0.
\end{align}
Imposing $y_2(0)=y_2'(0)=y_2(1)=y_2'(1)=0$ then gives
\begin{subequations}
\begin{align}
y_2(s) &= \frac{2b_{11}b_{20}}{3\lambda_\ast^3}\left\{\left[14 -\lambda_\ast^2(s - 2)\right]\sin{\lambda_\ast s} - \lambda_\ast\left[\lambda_\ast^2s(s - 1)(s - 2) + (2s + 3)\cos{\lambda_\ast s} + 12s - 3\right]\right\} + C_{y_2}y_1^{\text{(iii)}}(s),\\
\lambda_{-1} &= -\frac{2b_{11}b_{20}}{3a},
\end{align}
\end{subequations}
where $C_{y_2}$ is another constant of integration. Again, we need to continue the expansion. From \eqref{EL_complete y}, we now obtain
\begin{align}
    &12a^3\lambda_\ast^4\cos{2 \lambda_\ast s}+\frac{9a^3\lambda_\ast^4}{8} \bigl(\lambda_\ast^2-12\bigr) \cos{3 \lambda_\ast s}+3 a^3 \bigl(\lambda_\ast^2-4\bigr) \lambda_\ast^3 \sin{2 \lambda_\ast s}+\frac{9a^3\lambda_\ast^3 }{4} \bigl(4 - 3\lambda_\ast^2\bigr)  \sin{3\lambda_\ast s}\nonumber\\
    &\quad+\frac{\sin{\lambda_\ast s}}{36a\lambda_\ast}\left\{27 a^4 \lambda_\ast^6+32 b_{20} \lambda_\ast^2 \left[9 a^2 (b_{02}-2 b_{20})+3 a b_{11} C_{y_2}-b_{11}^2 b_{20} (s-2)\right]+36 a^2 \lambda_\ast^4 \bigl(3 a^2-2 b_{11}^2 s-2\bigr)+320 b_{11}^2 b_{20}^2\right\}\nonumber\\
    &\quad-\frac{\cos{\lambda_\ast s}}{72a}\left\{27 a^4 \lambda_\ast^6+36 a^2 \bigl(3 a^2-2\bigr) \lambda_\ast^4+96 a \lambda_\ast^2 \left[3 a \bigl(b_{20}b_{02}-2 b_{20}^2+b_{11}^2s\bigr)+b_{11} b_{20} C_{y_2}\right]+64 b_{11}^2 b_{20}^2 (2 s+1)\right\}\nonumber\\
    &\quad+\lambda_\ast^2\left\{4 a b_{02} b_{20}-4 a s \bigl(2 b_{02} b_{20}+b_{11}^2\bigr)+2(s - 1)\left[3b_{11}c_{30}(s + 1) + b_{20}c_{21} (s+3)\right]\right\}+\lambda_\ast^2 y_3''(s)+y_3''''(s) =0,
\end{align}
subject to $y_3(0)=y_3'(0)=y_3(1)=y_3'(1)=0$, which yields
\begin{subequations}
\begin{align}
    y_3(s) &= \frac{1}{576a\lambda_\ast^5}\Bigl[18 a^4 \bigl(3 \lambda_\ast^2-4\bigr) \lambda_\ast^4 \sin{3\lambda_\ast s}-9 a^4 \bigl(\lambda_\ast^2-12\bigr)\lambda_\ast^5 \cos{3\lambda_\ast s} - 144a^4 \bigl(\lambda_\ast^2-4\bigr) \lambda_\ast^4 \sin{2\lambda_\ast s}-576 a^4 \lambda_\ast^5 \cos{2 \lambda_\ast s}\nonumber\\
    &\hspace{17mm}- 16\sin{\lambda_\ast s}\Bigl(9a^4 \lambda_\ast^4\left[\lambda_\ast^2(2s - 1) - 16\right] + 2b_{11}^2 \left\{3a^2\lambda_\ast^2\left[\lambda_\ast^2 \bigl(6s^2 + 2s - 1\bigr) + 23\right]\right.\nonumber\\
    &\hspace{17mm}+\left.4 b_{20}^2 \left[\lambda_\ast^2\bigl(2s^2-6s-7\bigr)+21\right]\right\}+6 a b_{20} \left\{4 a b_{02} \lambda_\ast^2 \left[\lambda_\ast^2(1-2s)+16\right]+c_{21} \left[\lambda_\ast^4 (6s - 13)-100 \lambda_\ast^2 - 48\right]\right\}\nonumber\\ 
    &\hspace{17mm}+18ab_{11}c_{30}\left[\lambda_\ast^4 (2 s-5)-44 \lambda_\ast^2-48\right]\Bigr)+16 \lambda_\ast\cos{\lambda_\ast s}\Bigl(-6 a \lambda_\ast^2 (2 s+3) \bigl(3 a^3 \lambda_\ast^2-8 a b_{02} b_{20}+6 b_{20} c_{21}\bigr)\nonumber\\
    &\hspace{17mm}+b_{11}^2 \left\{3 a^2 \lambda_\ast^2 \bigl(6 \lambda_\ast^2 s^2-8 s+27\bigr)+4 b_{20}^2 \left[2 \lambda_\ast^2 (s-4) s-8 s+49\right]\right\}-36 a b_{11} c_{30} \lambda_\ast^2 (2 s+3)\Bigr)\nonumber\\
    &\hspace{17mm}+\lambda_\ast \Bigl(16 b_{11}^2 \left\{3 a^2 \lambda_\ast^2 \left[2 s\lambda_\ast^2 \bigl(4 s^2-1\bigr)+27(2s-1)\right]-4 b_{20}^2 \bigl(6 \lambda_\ast^2 s-50 s+49\bigr)\right\}-9a^4 \bigl(\lambda_\ast^2+148\bigr) \lambda_\ast^4 (2 s-1)\nonumber\\
    &\hspace{17mm}+96 a b_{20} \left\{4 a b_{02} \lambda_\ast^2 (2 s-1) \left[\lambda_\ast^2 (s-1) s+6\right]+c_{21} \left[2 \lambda_\ast^2 \bigl(6 s^2\!-\!44 s\!+\!9\bigr)-\lambda_\ast^4 (s-1)s\bigl(s^2\!+\!5s\!-\!13\bigr)-48 s\right]\right\}\nonumber\\
    &\hspace{17mm}+288 a b_{11} c_{30} \left[2 \lambda_\ast^2\bigl(6s^2-20s+3\bigr) - \lambda_\ast^4s\bigl(s^3-6 s+5\bigr)-48 s\right]\Bigr)\Bigr]+ C_{y_3}y_1^{\text{(iii)}}(s),\\
    C_{\lambda_0} &= \dfrac{27 a^4 \lambda_\ast^6+36 a^4 \lambda_\ast^4+24 a\lambda_\ast^2 \left[4 a b_{20} (5 b_{02}-6 b_{20})+13 a b_{11}^2+4 b_{11} b_{20}C_{y_2}-6 b_{11} c_{30}-6 b_{20} c_{21}\right]+416 b_{11}^2 b_{20}^2}{144 a^2\lambda_\ast^2}.
\end{align}
\end{subequations}
Again, we still need to determine $C_{y_2}$. Equation~\eqref{EL_complete lambda} gives
\begin{align}
    &\frac{2a(\lambda_\ast\sin{\lambda_\ast s} + 2\cos{\lambda_\ast s}-2)}{3\lambda_\ast^2}\Bigl(\lambda_\ast \sin{\lambda_\ast s}\left[3aC_{y_2}\lambda_\ast^2 + 4b_{11}b_{20} (s + 1)\right]-\lambda_\ast^2\left[6aC_{y_2}+b_{11} b_{20}\bigl(3s^2 + 4\bigr)\right]- 24b_{11} b_{20}\nonumber\\
    &+\cos{\lambda_\ast s}\left\{6aC_{y_2}\lambda_\ast^2+2 b_{11} b_{20} \left[12-\lambda_\ast^2(s - 2)\right]\right\}\Bigr)+\frac{4ab_{20}b_{11}s}{\lambda_\ast}\left[\bigl(\lambda_\ast^2s + 2\bigr)\sin{\lambda_\ast s} + \lambda_\ast (1-4s)+ \lambda_\ast(2s - 1)\cos{\lambda_\ast s} \right] \nonumber\\
    &+ 2x_3'(s)=0,
\end{align}
whence, requiring $x_3(0)=x_3(1)=0$,
\begin{subequations}
\begin{align}
    x_3(s) &= \frac{ab_{11}b_{20}}{6\lambda_\ast^2}\left\{3\lambda_\ast^2\left[s\bigl(4s^2 - 2s - 5\bigr) + 2\right] + 2\left[\lambda_\ast^2\bigl(3s^2 - 4\bigr) - 8(s - 2)\right]\cos{\lambda_\ast s}- \bigl(\lambda_\ast^2 - 4\bigr)(s - 2)\cos{2\lambda_\ast s}\right. \nonumber\\
    &\hspace{17mm}-4\lambda_\ast(s + 2)(3s - 4)\sin{\lambda_\ast s} + 4\lambda_\ast(s - 2)\sin{2\lambda_\ast s} + 12(s - 2)\bigr\},\\
    C_{y_2} &= -\frac{13b_{11}b_{20}}{3a\lambda_\ast^2}.
\end{align}
\end{subequations}
Finally, we can determine the buckling forces at $s=0$ and $s=1$, finding
\begin{subequations}
\begin{align}
    F_0 &= \frac{\lambda_\ast^2}{\ell^2} - \dfrac{1}{\ell}\left(\frac{4b_{11}b_{20}}{3a}\right) + \left[\frac{a^2\lambda_\ast^2}{8}\bigl(\lambda_\ast^2 - 4\bigr)-\frac{2(b_{11}c_{30} + b_{20}c_{21})}{a} + \frac{4b_{20}}{3} (5b_{02} - 6b_{20}) + \frac{13b_{11}^2}{3}\right] +  O(\ell),\\
    F_1&= F_0 + 2b_{20}^2\lambda_\ast^2+O(\ell).
\end{align}
\end{subequations}
Similarly to the symmetric case, this implies
\begin{align}
    F_1 &= \frac{\lambda_\ast^2}{\ell^2} - \dfrac{1}{\ell}\left(\frac{2\left|b_{11}b_{20}\right|\lambda_\ast}{3\sqrt{d}}\right) + \frac{1}{6}\biggl[3d \bigl(\lambda_\ast^2-4\bigr)+8b_{20}(5b_{02} - 6b_{20}) + 26b_{11}^2+2b_{20}^2\lambda_\ast^2 \nonumber\\
    &\hspace{70mm}-\left.\operatorname{sign}(b_{11}b_{20})\frac{6\lambda_\ast(b_{11}c_{30} + b_{20}c_{21})}{\sqrt{d}}\right] + O(\ell).\label{eq:F1asym}
\end{align}
Analogously to the discussion in the main text, this shows that, for all buckling modes, the expressions for the forces in \eqsref{eq:F1sym} and \neqref{eq:F1asym} lose asymptoticity when $d=O\bigl(\ell^{-2/3}\bigr)$ or $d=O\bigl(\ell^2\bigr)$. We are therefore left to discuss these cases of ``large'' and ``small'' buckling.

\subsubsection*{\textbf{``Large'' asymptotic buckling of a short relaxed elastic line}} We first consider the ``large'' buckling case, in which we write $\delta=\smash{\ell^{4/3}}d$ as discussed in the main text. The new buckling ansatz is therefore
\begin{subequations}
\begin{align}
    x(s) &= X(s) + \ell^{4/3} x_{4/3}(s) + \ell^2x_{2}(s) + O\bigl(\ell^{8/3}\bigr),\\
    y(s) &= Y(s) + \ell^{2/3} y_{2/3}(s) + \ell^{4/3}y_{4/3}(s) + \ell^2y_{2}(s) + O\bigl(\ell^{8/3}\bigr),\\
    \lambda(s) &=  \Lambda(s) + \ell^{-2}\lambda_{-2}(s) + \ell^{-4/3}\lambda_{-4/3}(s) + \ell^{-2/3}\lambda_{-2/3}(s) + \lambda_0(s) + O\bigl(\ell^{2/3}\bigr).
\end{align}
\end{subequations}
\paragraph*{\textbf{Leading-order solution.}} Again, we solve \eqsref{EL_complete} order-by-order. From \eqref{EL_complete x}, we find that $\smash{\lambda_{-2}'(s) = \lambda_{\smash{-4/3}}'(s) = 0}$. Setting $\lambda_{-2} = \lambda_\ast^2/2$ again, this turns \eqref{EL_complete y}, at leading order, into the eigenvalue problem
\begin{align}
\lambda_\ast^2y_{2/3}''(s)+y_{2/3}''''(s)=0,\quad\text{subject to }y_{2/3}(0)=y_{2/3}'(0)=y_{2/3}(1)=y_{2/3}'(1)=0.
\end{align}
This is the same problem that we solved earlier, so the eigenvalues and symmetric and asymmetric buckling modes are still given by \eqsref{eq:eigflat} and~\neqref{y1shapes}. 

\paragraph*{\textbf{Symmetric buckling.}} With $y_{2/3}(s)=y^{\text{(ii)}}_1(s)$, \eqsref{EL_complete} now give differential equations for $x_{4/3}(s)$, $\lambda_{-4/3}$, and $y_{4/3}(s)$, viz.,
\begin{align}
    &a^2\lambda_\ast^2\sin^2{\lambda_\ast s} + 2x_{4/3}'(s) = 0, &&\frac{3a^2\lambda_\ast^5}{2}\sin{2\lambda_\ast s} + 2\lambda_{-2/3}'(s)=0, &&2a\lambda_{-4/3}\lambda_\ast^2\cos{\lambda_\ast s} + \lambda_\ast^2y_{4/3}''(s) + y_{4/3}''''(s)=0.
\end{align}
Imposing $x_{4/3}(0)=y_{4/3}(0)=y'_{4/3}(0)=y_{4/3}(1)=y'_{4/3}(1)=0$, these equations yield 
\begin{align}
    &x_{4/3}(s) = -\frac{a^2\lambda_\ast}{8}(2\lambda_\ast s - \sin {2\lambda_\ast s}), &&y_{4/3}(s) = C_{y_{4/3}}\frac{1 - \cos{\lambda_\ast s}}{\lambda_\ast^2}, &&\lambda_{-4/3} = 0, &&\lambda_{-2/3}(s) = C_{\lambda_{-2/3}} + \frac{3a^2\lambda_\ast^4}{8}\cos{2\lambda_\ast s},
\end{align}
where $C_{y_{4/3}},C_{\lambda_{-2/3}}$ are constants of integration that remain undetermined at this stage. Requiring $x_{4/3}(1)=-d$ recovers the now familiar relation~\neqref{eq:a} between the buckling amplitude $a$ and the compression $d$. We continue the expansion using \eqref{EL_complete y}, which now yields
\begin{align}
    \frac{\lambda_\ast^2}{8}\left[9a^3\lambda_\ast^4 \cos{3\lambda_\ast s} + a\bigl(16C_{\lambda_{-2/3}} - 3a^2\lambda_\ast^4\bigr) \cos{\lambda_\ast s} + 32b_{11}b_{20}(s - 1)\right] + \lambda_\ast^2 y_2''(s) + y_2''''(s)=0,
\end{align}
subject to $y_2(0)=y_2'(0)=y_2(1)=y_2'(1)=0$. This implies
\begin{subequations}
\begin{align}
    y_2(s) &= \frac{1}{192 \lambda_\ast^2}\left\{\lambda_\ast\sin{\lambda_\ast s}\left[3a^3 \lambda_\ast^2+128 b_{11} b_{20} (2-3 s)\right]-3a^3 \lambda_\ast^4 (s-1)-3a^3 \lambda_\ast^4 \cos{3\lambda_\ast s}+3 a^3 \lambda_\ast^3 \cos{3 \lambda_\ast} (\lambda_\ast s-\sin{\lambda_\ast s})\right.\nonumber\\
    &\hspace{16mm}+\left.64 \lambda_\ast^2\left[3aC_{y_2}-2b_{11}b_{20} (s-2) (s-1) s\right]-192 \cos{\lambda_\ast s} \bigl(aC_{y_2} \lambda_\ast^2+6b_{11} b_{20}\bigr)+1152b_{11}b_{20}\right\},\\
    C_{\lambda_{-2/3}} &=\frac{3a^2\lambda_\ast^4}{16} - \frac{2b_{11}b_{20}}{a},
\end{align}
\end{subequations}
in which $C_{y_2}$ is another constant of integration. From~\eqref{EL_complete lambda}, 
\begin{align}
    aC_{y_{4/3}}\sin^2{\lambda_\ast s} + x_2'(s) = 0 \quad\Longrightarrow \quad x_2(s) = -\frac{aC_{y_{4/3}}(2\lambda_\ast s - \sin{2\lambda_\ast s})}{4\lambda_\ast},
\end{align}
where we have imposed $x_2(0)=0$. The condition $x_2(1) = 0$ now implies $C_{y_{4/3}} =0$, whence $x_2(s) = y_{4/3}(s) = 0$. We can now compute the buckling forces at $s=0$ and $s=1$,
\begin{align}
    &F_0 = \frac{\lambda_\ast^2}{\ell^2} + \dfrac{1}{\ell^{2/3}}\left(\dfrac{a^2\lambda_\ast^4}{8}-\dfrac{4b_{11}b_{20}}{a}\right) + O(1)=F_1+O(1).
\end{align}
As above, this implies that, in the lowest buckling mode, $a=a_\pm$ if $b_{11}b_{20}\gtrless0$. Now, in particular, 
\begin{align}
F_1^\pm = \frac{\lambda_\ast^2}{\ell^2} + \dfrac{1}{\ell^{2/3}}\left(\dfrac{d\lambda_\ast^2}{2}\mp\dfrac{2\lambda_\ast\left|b_{11}b_{20}\right|}{\sqrt{d}}\right) + O(1).
\end{align}
In the lowest buckling mode, $\lambda_\ast=2\pi$, so, on identifying $b_{20}=A$, $b_{11}=B$, we recover \eqref{eq:largeb} of the main text, as required. More generally, this shows that, for each buckling mode, one of ``up'' or ``down'' buckling undergoes a snap-through instability, as discussed in the main text for the lowest mode.
\paragraph*{\textbf{Asymmetric buckling.}} Now taking $y_{2/3}(s)=y^{\text{(iii)}}_1(s)$, \eqsref{EL_complete} again give differential equations for $x_{4/3}(s)$, $\lambda_{-4/3}$, and~$y_{2/3}$, \begin{subequations}
\begin{align}
    a^2(\lambda_\ast\sin{\lambda_\ast s} + 2\cos{\lambda_\ast s} - 2)^2 + 2x_{(4/3)}'(s) &= 0,\\
    \frac{3a^2\lambda_\ast^3}{2}\left[\bigl(\lambda_\ast^2 - 4\bigr)\sin{2\lambda_\ast s} + 4\lambda_\ast\cos{2\lambda_\ast s}\right] + 2\lambda_{-2/3}'(s) &= 0, \\
    2a\lambda_\ast\lambda_{-4/3} (\lambda_\ast\cos{\lambda_\ast s} - 2\sin{\lambda_\ast s}) + \lambda_\ast^2 y_{4/3}''(s) + y_{4/3}''''(s)&= 0,
\end{align}
\end{subequations}
Imposing $x_{4/3}(0)=y_{4/3}(0)=y'_{4/3}(0)=y_{4/3}(1)=y'_{4/3}(1)=0$ again, these imply\begin{subequations}
\begin{align}
    x_{4/3}(s) &= \frac{a^2}{8\lambda_\ast}\left[\bigl(\lambda_\ast^2 - 4\bigr)\sin{2\lambda_\ast s} + 32\sin{\lambda_\ast s}- 16\lambda_\ast\cos{\lambda_\ast s} + 4\lambda_\ast\cos{2\lambda_\ast s}-2\lambda_\ast\bigl(\lambda_\ast^2 s+ 12s - 6\bigr)\right],\\
    y_{4/3}(s) &= \frac{C_{y_{4/3}}}{\lambda_\ast^3}\bigl(2\lambda_\ast s - \lambda_\ast - 2\sin{\lambda_\ast s} + \lambda_\ast\cos{\lambda_\ast s}\bigr),\\
    \lambda_{-4/3} &= 0,\qquad\lambda_{-2/3} = C_{\lambda_{-2/3}} + \frac{3a^2\lambda_\ast^2}{8}\left[\bigl(\lambda_\ast^2 - 4\bigr)\cos{2\lambda_\ast s} - 4\lambda_\ast\sin{2\lambda_\ast s}\right],
\end{align}
\end{subequations}
with the constants of integration $C_{y_{4/3}},C_{\lambda_{-2/3}}$ still to be determined. The expansion of \eqref{EL_complete y} now gives
\begin{align}
    &\frac{a\lambda_\ast}{4}\bigl(\lambda_\ast\cos{\lambda_\ast s} - 2 \sin{\lambda_\ast s}\bigr)\left\{3a^2\lambda_\ast^2\left[3\bigl(\lambda_\ast^2 - 4\bigr) \cos{2\lambda_\ast s} + 8\lambda_\ast\sin{\lambda_\ast s} - 12 \lambda_\ast\sin{2\lambda_\ast s} + 16\cos{\lambda_\ast s}-6a^2\bigl(\lambda_\ast^2 + 4\bigr) \lambda_\ast^2\right]\right.\nonumber\\
    &\hspace{45mm}+8C_{\lambda_{-2/3}}\bigr\} + 4b_{11}b_{20}\lambda_\ast^2(s - 1) + \lambda_\ast^2 y_2''(s) + y_2''''(s)=0,
\end{align}
subject to $y_2(0)=y_2'(0)=y_2(1)=y_2'(1)=0$. Introducing another constant of integration $C_{y_2}$, we find
\begin{subequations}
\begin{align}
    y_2(s) &= \frac{1}{192\lambda_\ast^3}\Bigl(\lambda_\ast^2\left\{-16 \sin{\lambda_\ast s}\left[a^3 \bigl(6\lambda_\ast^2s - 3\lambda_\ast^2 - 48\bigr)-24 a C_{y_2}+8 b_{11} b_{20} (s-2)\right]-48 a^3 \bigl(\lambda_\ast^2-4\bigr) \sin{2 \lambda_\ast s}\right.\nonumber\\
    &\hspace{16mm}+\left.6 a^3 \bigl(3 \lambda_\ast^2-4\bigr) \sin{3 \lambda_\ast s}-3 a^3 \lambda_\ast \bigl(\lambda_\ast^2-12\bigr) \cos{3 \lambda_\ast s}-192 a^3 \lambda_\ast \cos{2 \lambda_\ast s}\right\} +1792 b_{11} b_{20} \sin{\lambda_\ast s}\nonumber\\
    &\hspace{16mm} +128 b_{11} b_{20} \lambda_\ast \left[3-\lambda_\ast^2s(s - 2)(s - 1)-12 s\right] - 3a \lambda_\ast^3 (2 s-1) \left[a^2 \bigl(\lambda_\ast^2+148\bigr)+64 C_{y_2}\right]\nonumber\\
    &\hspace{16mm}-32 \lambda_\ast \cos{\lambda_\ast s} \left[3a\lambda_\ast^2\bigl(2a^2s + 3a^2 + 2C_{y_2}\bigr) + 4b_{11}b_{20}(2s + 3)\right]\Bigr),\\
    C_{\lambda_{-2/3}} &= \frac{a^2\lambda_\ast^2}{16}\bigl(3\lambda_\ast^2 + 4\bigr) - \frac{2b_{11}b_{20}}{3a}. 
\end{align}
\end{subequations}
From \eqref{EL_complete lambda}, we now obtain an equation for $x_2(s)$,
\begin{align}
    4aC_{y_{4/3}}\sin^2{\frac{\lambda_\ast s}{2}} \left(\lambda_\ast \cos{\frac{\lambda_\ast s}{2}}-2 \sin{\frac{\lambda_\ast s}{2}}\right)^2+\lambda_\ast^2 x_2'(s)=0,
\end{align}
which, on imposing $x_2(0)=0$, integrates to
\begin{align}
    x_2(s) = \frac{aC_{y_{4/3}}}{4\lambda_\ast^3}\left[-2\lambda_\ast \bigl(\lambda_\ast^2+12\bigr) s+\bigl(\lambda_\ast^2-4\bigr) \sin{2\lambda_\ast s}+32\lambda_\ast \sin ^4{\dfrac{\lambda_\ast s}{2}}+32 \sin{\lambda_\ast s}\right],
\end{align}
However, the boundary condition $x_2(1) = 0$ implies $C_{y_{4/3}} =0$, whence $x_2(s) = y_{4/3}(s) = 0$. We can now obtain the buckling forces at $s=0$ and $s=1$,
\begin{align}
    &F_0 = \frac{4\pi^2}{\ell^2} + \dfrac{1}{\ell^{2/3}}\left[\frac{a^2\lambda_\ast^2}{8}\bigl(\lambda_\ast^2-4\bigr)-\frac{4 b_{11}b_{20}}{3a}\right] + O(1)=F_1 + O(1).
\end{align}
Again, this shows that $a=a_\pm$ if $b_{11}b_{20}\gtrless 0$ in the lowest buckling mode. More generally, this shows that, as for the symmetric buckling mode, on of ``up'' and ``down'' buckling undergoes a snap-through instability in each asymmetric mode.

\subsubsection*{\textbf{``Small'' asymptotic buckling of a short relaxed elastic line}} We now turn to the ``small'' buckling case, in which we write $\delta=\ell^4d$, as discussed in the main text. The new buckling ansatz is therefore
\begin{align}
    &x(s) = X(s) + \ell^{4} x_4(s) + O\bigl(\ell^5\bigr),
    &&y(s) = Y(s) + \ell^2 y_2(s) + O\bigl(\ell^3\bigr),
    &&\lambda(s) =  \Lambda(s) + \dfrac{\lambda_{-2}(s)}{\ell^2} + \dfrac{\lambda_{-1}(s)}{\ell} + O(1),
\end{align}
so boundary conditions~\neqref{BC_buckling} imply, in particular $x_4(1) = -d$. On expanding \eqref{EL_complete x}, we find that $\lambda_{-2}$ and $\lambda_{-1}$ are both constants. Again, we write $\lambda_{-2} = \lambda_\ast^2/2$. Equation~\neqref{EL_complete y} then gives
\begin{align}
    4b_{11}b_{20}\lambda_\ast^2(s-1)+\lambda_\ast^2 y_2''(s)+y_2''''(s) = 0\quad\text{subject to }y_2(0)=y_2'(0)=y_2(1)=y_2'(1)=0.
\end{align}
This means that even the leading-order buckling problem now differs from the corresponding problem in the flat case if and only if $b_{11}b_{20}\neq 0$. This heralds the fundamental differences between flat buckling and buckling on general curved surfaces. If $b_{11}b_{20}\neq 0$, the solution of this problem is
\begin{align}
    y_2(s) &= -\frac{2b_{11}b_{20}}{3\lambda_\ast(\lambda_\ast\sin{\lambda_\ast} + 2\cos{\lambda_\ast} - 2)}\left\{\lambda_\ast\bigl(1-3s+6s^2-2s^3\bigr)+\left[\lambda_\ast^2s(s-1)(s-2)-3\right]\sin{\lambda_\ast} + 3\sin{\lambda_\ast s}- \lambda_\ast\cos {\lambda_\ast s}\right.\nonumber\\
    &\hspace{51mm}+\left. 3\sin{[\lambda_\ast(1-s)]} - 2\lambda_\ast\cos{[\lambda_\ast(1-s)]} + \lambda_\ast(s - 2)\bigl(2s^2 - 2s - 1\bigr)\cos{\lambda_\ast}\right\}.
\end{align}
In particular $\lambda_\ast$ now remains undetermined at leading order. Equation~\neqref{EL_complete lambda} now yields a differential equation for $x_4(s)$,
\begin{align}
    &\dfrac{1}{3}\left(\frac{\lambda_\ast \left\{\left(3s^2+4\right)\sin{\lambda_\ast} + 2\sin{\lambda_\ast s} - 4\sin{[\lambda_\ast(1-s)]}\right\} - 6\left\{\left(s^2 + 1\right)(1-\cos{\lambda_\ast})+\cos{[\lambda_\ast(1-s)]} + \cos{\lambda_\ast s}\right\}}{\lambda_\ast\sin{\lambda_\ast} + 2\cos{\lambda_\ast} - 2}\right)^2\nonumber\\
    &\quad-\frac{8s}{\lambda_\ast(\lambda_\ast\sin{\lambda_\ast} + 2\cos{\lambda_\ast} - 2)}\left\{\lambda_\ast\bigl(1-6s+18s^2-8s^3\bigr)+\left[\lambda_\ast^2s\bigl(4s^2 - 9s + 4\bigr)-3\right] \sin{\lambda_\ast}+\bigl(\lambda_\ast^2s+3\bigr)\sin{\lambda_\ast s}\right.\nonumber\\
    &\qquad\qquad+\left.\bigl(3-2\lambda_\ast^2s\bigr)\sin{[\lambda_\ast(1-s)]}  + 2\lambda_\ast\bigl(4s^3 - 9s^2 + 3s + 1\bigr) \cos{\lambda_\ast}+ \lambda_\ast(3s - 1)\cos{\lambda_\ast s} -\lambda_\ast(3s+2)\cos{[\lambda_\ast(1-s)]}\right\}\nonumber\\
    &\quad-3(s - 4)^2s^2 + \dfrac{6x_4'(s)}{b_{11}^2b_{20}^2} = 0,
\end{align}
which gives, on imposing $x_4(0)=0$,
\begin{align}
    x_4(s) = &-\dfrac{b_{11}^2b_{20}^2 \csc^4\dfrac{\lambda_\ast}{2}}{360\lambda_\ast\left(\lambda_\ast\cot\dfrac{\lambda_\ast}{2} - 2\right)^2}\Bigl(60\lambda_\ast^2\sin{\lambda_\ast}+60 \lambda_\ast^2 \sin{2\lambda_\ast}+384 \lambda_\ast^2 s^5 \sin{\lambda_\ast}-192 \lambda_\ast^2 s^5 \sin{2 \lambda_\ast}- 1440\lambda_\ast^2s^4\sin{\lambda_\ast}\nonumber\\
    &\qquad + 720\lambda_\ast^2s^4 \sin{2\lambda_\ast} + 1380\lambda_\ast^2s^3\sin{\lambda_\ast} - 690\lambda_\ast^2s^3\sin{2\lambda_\ast} - 3\left\{-90\sin{\lambda_\ast}+ 2\lambda_\ast^3s\bigl(8s^4-30s^3+30s^2-15\bigr)\right.\nonumber\\
    &\qquad+45 \sin{2\lambda_\ast} + 6\lambda_\ast\bigl(32s^5-120s^4+110s^3- 5s^2 - 25s + 5\bigr)+ 60\sin{[\lambda_\ast(s - 2)]} + 180\sin{\lambda_\ast s} - 15\sin{2 \lambda_\ast s}\nonumber\\
    &\qquad -60\sin{[\lambda_\ast(s +1)]} + 180\sin{[\lambda_\ast(1-s)]}-30\sin{[\lambda_\ast(1-2s)]} + 15\sin{[2\lambda_\ast(1-s)]}\bigr\} - 60\lambda_\ast^2s^2\sin{\lambda_\ast}\nonumber\\
    &\qquad- 60\lambda_\ast^2s^2\sin{2\lambda_\ast} - 60\lambda_\ast^2s^2\sin{[\lambda_\ast (s - 2)]}+ 60\lambda_\ast^2s^2\sin{\lambda_\ast s} + 10 \bigl(3\lambda_\ast^2s^2 - 18s^2 - 4\lambda_\ast^2\bigr)\sin{[\lambda_\ast(s +1)]}\nonumber\\
    &\qquad+ 10 \bigl(3\lambda_\ast^2s^2 + 54s^2 - 4\lambda_\ast^2\bigr)\sin{[\lambda_\ast(1-s)]}- 360s^2\sin{\lambda_\ast} + 180s^2\sin{2\lambda_\ast} + 180s^2\sin{[\lambda_\ast(s - 2)]}\nonumber\\
    &\qquad+ 2\lambda_\ast\cos{2\lambda_\ast}\left[2\lambda_\ast^2s\bigl(12s^4-45s^3+45s^2-10\bigr) - 96s^5+360s^4-330s^3-105s^2+45s+ 90\right]\nonumber\\
    &\qquad+ 540s^2\sin{\lambda_\ast s} + 210\lambda_\ast s^2\cos{[\lambda_\ast(s - 2)]}- 90\lambda_\ast s^2\cos{\lambda_\ast s} - 270\lambda_\ast s^2\cos{[\lambda_\ast(1-s)]}- 420\lambda_\ast^2s\sin{\lambda_\ast} \nonumber\\ 
    &\qquad+ 120\lambda_\ast^2s\sin{2\lambda_\ast} + 80\lambda_\ast^2\sin{[\lambda_\ast (s - 2)]} - 80\lambda_\ast^2\sin{\lambda_\ast s} - 5\lambda_\ast^2\sin{2\lambda_\ast s} + 20\lambda_\ast^2\sin{[\lambda_\ast(1-2s)]}\nonumber\\
    &\qquad+ 20\lambda_\ast^2\sin{[2\lambda_\ast(1-s)]} + 2\lambda_\ast\cos{\lambda_\ast}\left[384s^5-1440s^4+1320s^3+60s^2+10\bigl(2\lambda_\ast^2-27\bigr)s-45\right]\nonumber\\
    &\qquad- 240\lambda_\ast\cos{[\lambda_\ast (s - 2)]} + 60\lambda_\ast\cos{[2\lambda_\ast(s-1)]} + 120\lambda_\ast\cos{\lambda_\ast s} - 30\lambda_\ast\cos{2\lambda_\ast s}+ 300 \lambda_\ast\cos{[\lambda_\ast(1-s)]}\nonumber\\
    &\qquad- 30 \lambda_\ast\cos{[\lambda_\ast(1-2s)]}+ 30\lambda_\ast\bigl(5s^2 - 6\bigr)\cos{[\lambda_\ast(s+1)]}\Bigr).
\end{align}
The boundary condition $x_4(1)=-d$ then yields a relation between $d$ and $\lambda_\ast$,
\begin{align}
    d &= \dfrac{b_{11}^2b_{20}^2\left\{2\lambda_\ast \left[21\lambda_\ast^2+72 + \bigl(20\lambda_\ast^2 - 6\bigr)\cos{\lambda_\ast} + \bigl(4 \lambda_\ast^2 - 66\bigr)\cos{2\lambda_\ast}\right]- 3\left[\bigl(52\lambda_\ast^2+60\bigr)\sin{\lambda_\ast}+\left(19\lambda_\ast^2-30\right)\sin{2\lambda_\ast}\right]\right\}}{90 \lambda_\ast \left(\lambda_\ast\sin{\lambda_\ast}+2\cos{\lambda_\ast}-2\right)^2},
\end{align}
which is \eqref{eq:l} in the main text. In particular, $d\to\infty$ as $\lambda_\ast\to2\pi$, which is consistent with recovering the result for ``large'' buckling. Conversely, as noted in the main text,
\begin{align}
    \lambda_\ast= \frac{175d}{b_{11}^2b_{20}^2} +O\bigl(d^2\bigr).
\end{align}
We further obtain $F_0=\ell^{-2}\lambda_\ast^2+O\bigl(\ell^{-1}\bigr)=F_1$. This is \emph{consistent} with $F_\ast=0$, but only proves that $F_\ast=O\bigl(\ell^{-1}\bigr)$, i.e., is asymptotically smaller than in the flat case. Higher-order terms could of course be obtained by continuing this expansion, but proving that $F_\ast=0$ would require knowledge of all higher-order terms. Obtaining this terms or proving that $F_\ast=0$ in a different way is beyond the scope of our calculations.

\subsection{Intrinsic Euler buckling on curved surfaces}
As discussed in the main text, the previous section analysis ``extrinsic'' buckling because the Lagrangian minimises the total curvature of the elastic line, which is an extrinsic quantity. Intrinsic buckling replaces \eqref{eq:L} of the main text with
\begin{align}
    \mathcal{L} =\frac{1}{2} \int_0^1{\dfrac{\kappa_\text{g}(s)^2}{\ell^2}\,\mathrm{d}s} - \int_0^1\lambda(s)\left[\alpha(s)^2 - 1\right]\mathrm{d}s,\label{eq:Lint}
\end{align}
where $\kappa_\text{g}(s)$ is the geodesic curvature of the elastic line. Let $\vec{r}(x,y)=\bigl(x,y,h(x,y)\bigr)$ be a point on the surface to which the elastic line is confined, where $h(x,y)$ is still given by \eqref{eq:h}. The coefficients of its first fundamental form~\cite{docarmo} are therefore
\begin{align}
&E(x,y)=\dfrac{\partial\vec{r}}{\partial x}\cdot\dfrac{\partial\vec{r}}{\partial x},&&F(x,y)=\dfrac{\partial\vec{r}}{\partial x}\cdot\dfrac{\partial\vec{r}}{\partial y},&&G(x,y)=\dfrac{\partial\vec{r}}{\partial y}\cdot\dfrac{\partial\vec{r}}{\partial y}.
\end{align}
In terms of these coefficients,
\begin{subequations}
\begin{align}
\alpha(s)^2&=E\bigl(x(s),y(s)\bigr)x'(s)^2+2F\bigl(x(s),y(s)\bigr)x'(s)y'(s)+G\bigl(x(s),y(s)\bigr)y'(s)^2,\\
\kappa_\text{g}(s)&=\sqrt{g\bigl(x(s),y(s)\bigr)}\left\{{\Gamma^y}_{xx}\bigl(x(s),y(s)\bigr)x'(s)^3\!-\!{\Gamma^x}_{yy}\bigl(x(s),y(s)\bigr)y'(s)^3\!+\!\left[2{\Gamma^y}_{xy}\bigl(x(s),y(s)\bigr)\!-\!{\Gamma^x}_{xx}\bigl(x(s),y(s)\bigr)\right]x'(s)^2y'(s)\right.\nonumber\\
&\hspace{30mm}-\left.\left[2{\Gamma^x}_{xy}\bigl(x(s),y(s)\bigr)-{\Gamma^y}_{yy}\bigl(x(s),y(s)\bigr)\right]x'(s)y'(s)^2 + x'(s)y''(s)-x''(s)y'(s)\right\},
\end{align}
\end{subequations}
in which $g=EG-F^2$ and the Christoffel symbols satisfy
\begin{subequations}
\begin{align}
    &{\Gamma^x}_{xx} = \frac{GE_{,x} - 2FF_{,x} + FE_{,y}}{2g},
    &&{\Gamma^y}_{xx} = \frac{2EF_{,x} - EE_{,y} - FE_{,x}}{2g},
    &&{\Gamma^x}_{xy} = \frac{GE_{,y}-FG_{,x}}{2g},\\
    &{\Gamma^y}_{yy} = \frac{EG_{,y} - 2FF_{,y} + FG_{,x}}{2g},
    &&{\Gamma^x}_{yy} = \frac{2GF_{,y} - GG_{,x} - FG_{,y}}{2g},
    &&{\Gamma^y}_{xy} = \frac{EG_{,x} - FE_{,y}}{2g},
\end{align}
\end{subequations}
where commata denote differentiation. Meanwhile, the boundary conditions of the problem remain unchanged. To solve the buckling problem, we therefore need to determine, again, first the shape of the relaxed line, and then compute the buckled shapes under compression. It is clear that \eqref{eq:Lint} is minimised if $\kappa_\text{g}(s)\equiv 0$, and this minimum value is attained by a geodesic.

\subsubsection*{\textbf{Asymptotic calculation of the shape of a geodesic}}
To determine the shape of the relaxed elastic line for $\ell\ll1$, we could again solve the Euler--Lagrange equations, but it is more elegant to solve the geodesic equations asymptotically. They are~\cite{docarmo}
\begin{subequations}\label{geodesic equations}
\begin{align}
\label{geodesic equation1}
    x''(s) + {\Gamma^x}_{xx}\bigl(x(s),y(s)\bigr) x'(s)^2 +2{\Gamma^x}_{xy}\bigl(x(s),y(s)\bigr)x'(s)y'(s)+{\Gamma^x}_{yy}\bigl(x(s),y(s)\bigr)y'(s)^2=0,\\
\label{geodesic equation2}
    y''(s) + {\Gamma^y}_{xx}\bigl(x(s),y(s)\bigr) x'(s)^2 +2 {\Gamma^y}_{xy}\bigl(x(s),y(s)\bigr)x'(s)y'(s)+{\Gamma^y}_{yy}\bigl(x(s),y(s)\bigr)y'(s)^2=0.
\end{align}
\end{subequations}
Again assuming the geodesic to be clamped at the origin in the $x$-direction without loss of generality, the boundary conditions for these equations are
\begin{align}
\label{geodesic BC}
    &x(0)=0,&&y(0)=0,&&x'(0)=1,&&y'(0)=0.
\end{align}
Similarly to the extrinsic problem, we seek the solution of \eqsref{geodesic equations} subject to boundary conditions \neqref{geodesic BC} in the form
\begin{align}
    x(s) &= x_0(s) + \ell x_1(s) + \ell^2 x_2(s) + \ell^3 x_3(s) +  O\bigl(\ell^4\bigr),&y(s) &= y_0(s) + \ell y_1(s) + \ell^2 y_2(s) + \ell^3 y_3(s) + O\bigl(\ell^4\bigr).
\end{align}
\paragraph*{\textbf{Solution at orders $\vec{O(1)}$ and $\vec{O(\ell)}$.}} The geodesic equations~\eqref{geodesic equations} yield
\begin{align}
    &x_0''(s) =0,&&y_0''(s)=0,&&x_1''(s) =0, &&y_1''(s)=0,
\end{align}
subject to $x_0(0)=y_0(0)=y_0'(0)=x_1(0)=x_1'(0)=y_1(0)=y_1'(0)=0$ and $x_0'(1)=1$, so 
\begin{align}
    &x_0(s) = s, &&y_0(s) = 0,&&x_1(s) = 0, &&y_1(s) = 0.
\end{align}
\paragraph*{\textbf{Solution at orders $\vec{O(\ell^2)}$ and $\vec{O(\ell^3)}$.}} The geodesic equations~\eqref{geodesic equations} now give
\begin{align}
    &4 b_{20}^2 s + x_2''(s) = 0, &&18 b_{20} c_{30} s^2 + x_3''(s) =0, &&2 b_{20}b_{11} s + y_2''(s) = 0, &&2 b_{20} c_{21} s^2 + 6b_{11}c_{30}s^2 + y_3''(s) = 0,
\end{align}
subject to $x_2(0)=x_2'(0)=y_2(0)=y_2'(0)=x_3(0)=x_3'(0)=y_3(0)=y_3'(0)=0$. Hence
\begin{align}
    &x_2(s) = -\frac{2b_{20} s^3}{3}, &&x_3(s) = -\frac{3b_{20}c_{30}s^4}{2}, &&y_2(s) = -\frac{b_{20}b_{11} s^3}{3}, &&y_3(s) = -\frac{s^4}{6}(b_{20}c_{21}+3b_{11}c_{30}).
\end{align}
Combining these solutions, the geodesic, i.e. the shape of the relaxed elastic line is now
\begin{align}
    &X(s) = s -\frac{2b_{20} s^3}{3} \ell^2 -\frac{3b_{20}c_{30}s^4}{2}\ell^3 + O\bigl(\ell^4\bigr), 
    &&Y(s) = -\frac{b_{20}b_{11} s^3}{3} \ell^2 -\frac{s^4}{6}(b_{20}c_{21}+3b_{11}c_{30})\ell^3 + O\bigl(\ell^4\bigr).\label{eq:geodesic}
\end{align}

\subsubsection*{\textbf{Asymptotic buckling of a short geodesic}}
We are now ready to solve the buckling problem: We solve the Euler--Lagrange equations of \eqref{eq:Lint} subject to the boundary conditions~\neqref{BC_buckling}, noting that the geodesic shape~\neqref{eq:geodesic} replaces the previous result~\neqref{relaxed elastic line} for the shape of the relaxed elastic line. Again, we assume that the relative compression has the scaling $\delta=\ell^2d$, with $d=O(1)$, and take the buckling ansatz
\begin{align}
    &x(s) = X(s) + \ell^{2} x_{2}(s) + O\bigl(\ell^{4}\bigr),
    &&y(s) = Y(s) + \ell y_{1}(s) + \ell^{3}y_{3}(s) + O\bigl(\ell^{5}\bigr),
    &&\lambda(s) =  \ell^{-2}\lambda_{-2}(s) + \lambda_{0}(s) +O\bigl(\ell^{2}\bigr),
\end{align}
We note that this ansatz is simpler than the ansatz~\neqref{eq:ansatz} that we used for the extrinsic buckling problem: For example, there is no term at order $O\bigl(\ell^2\bigr)$ in the expansion of $y(s)$. The fact that this simpler ansatz works is the reason why the intrinsic buckling problem will turn out to be less interesting. 

\paragraph*{\textbf{Leading-order solution.}} At leading order, we find that $\lambda_{-2}(s)=\lambda_\ast^2/2$ is constant, from which we obtain, again, the eigenvalue problem $\lambda_\ast^2 y_1''(s) + y_1''''(s) = 0$, subject to $y_1(0)=y_1'(0)=y_1(1)=y_1'(1)=0$. This means that the eigenvalues are still given by \eqsref{eq:lflat}, with symmetric and asymmetric buckling modes given by \eqsref{y1shapes}. 

\paragraph*{\textbf{Symmetric buckling.}} As in the case of extrinsic buckling, we first analyse the symmetric buckling modes. With $\smash{y_1(s)=y_1^{\text{(ii)}}(s)}$, and expanding the Euler--Lagrange equations of \eqref{eq:Lint}, we find
\begin{align}
    &2\lambda_0'(s) + \frac{5a^2\lambda_\ast^5}{2}\sin{2\lambda_\ast s}= 0, &&2x_2'(s) +a\lambda_\ast^2\sin^2{\lambda_\ast s} = 0,
\end{align}
which, with $x_2(0)=0$, integrate to
\begin{align}
    &\lambda_0(s) = C_{\lambda_0} + \frac{5a^2\lambda_\ast^4}{8}\cos{2\lambda_\ast s}, &&x_2(s) = -\frac{a^2\lambda_\ast}{8}(2\lambda_\ast s - \sin{2\lambda_\ast s}),
\end{align}
where $C_{\lambda_0}$ is a constant of integration. The boundary condition $x_2(1)=-d$ shows that the relation \neqref{eq:a} between the buckling amplitude $a$ and $d$ continues to hold. At next order, we find
\begin{align}
    &\frac{a\lambda_\ast^2}{8}\left\{\cos{\lambda_\ast s} \left[16(2b_{02}b_{20} + 2b_{11}^2 + C_{\lambda_0}) - 7a^2\lambda_\ast^4\right]+ 9a^2\lambda_\ast^4\cos{3\lambda_\ast s} + 32b_{02}b_{20} - 16b_{11}^2\lambda_\ast s\sin{\lambda_\ast s}\right\} + \lambda_\ast^2y_3''(s) + y_3''''(s) = 0.
\end{align}
Imposing $y_3(0)= y_3'(0)= y_3(1)= y_3'(1)= 0$, this implies
\begin{subequations}
\begin{align}
    y_3(s) &= \frac{1}{64\lambda_\ast^2}\left\{a^3\lambda_\ast^4 - a^3\lambda_\ast^4\cos{3\lambda_\ast s} + 16a\cos{\lambda_\ast s}\left[48b_{02}b_{20} + b_{11}^2\bigl(2\lambda_\ast^2s^2 + 1\bigr) \right] - 16a\bigl(48b_{02}b_{20} +  b_{11}^2\bigr)\right. \nonumber\\
    &\hspace{14mm}\left.- 32a\lambda_\ast^2s\left[4b_{02}b_{20}(s - 1) + b_{11}^2\right] - 32a\lambda_\ast(2s - 1)(b_{11}^2 - 4b_{02}b_{20})\sin{\lambda_\ast s}\right\}+C_{y_3}y_1^{\text{(ii)}}(s), \\ 
    C_{\lambda_0} &= \frac{7a^2\lambda_\ast^4}{16}+2b_{02}b_{20} - \frac{b_{11}^2}{2},
\end{align}
\end{subequations}
in which $C_{y_3}$ is another constant of integration that do not need to determine. Indeed, we can now obtain the buckling forces at $s=0$ and $s=1$,
\begin{align}
    &F_0 =\frac{\lambda_\ast^2}{\ell ^2} + \frac{a^2\lambda_\ast^4}{8} + 4b_{02}b_{20}-b_{11}^2 + O(\ell) = \frac{\lambda_\ast^2}{\ell ^2} + \frac{d\lambda_\ast^2}{2} + K + O(\ell), &&F_1 = F_0 + 2b_{20}^2\lambda_\ast^2 + O(\ell),
\end{align}
using \eqref{eq:a} and expressing the result in terms of the Gaussian curvature $K= 4b_{02}b_{20}-b_{11}^2$ of the surface at the origin. Unlike extrinsic buckling, ``up'' and ``down'' buckling are equivalent, even if $b_{11}b_{20}\neq 0$. Moreover, for each buckling mode, the associated critical force is $F_\ast=\lambda_\ast^2/\ell^2+K\,\big(\!+\,2b_{20}^2\lambda_\ast^2\big)$, which, by comparison with the results for flat buckling, is seen to represent a mere shift of the critical force in that case. Finally, for the lowest buckling mode, $\lambda_\ast=2\pi$, and identifying $b_{20}=A$, $b_{11}=B$, we recover \eqref{eq:Fint} of the main text, as required.

\paragraph*{\textbf{Asymmetric buckling.}} Finally, we analyse asymmetric intrinsic buckling. Once again, we will make use of \eqsref{eq:asymsimp} to simplify the results. Taking $\smash{y_1(s)=y_1^{\text{(iii)}}(s)}$, the Euler--Lagrange equations of \eqref{eq:Lint} yield
\begin{align}
    &2\lambda_0'(s) + \frac{5a^2\lambda_\ast^3}{2} \left[(\lambda_\ast^2 - 4)\sin{2\lambda_\ast s} + 4\lambda_\ast\cos{2\lambda_\ast s}\right] = 0, 
    &&2x_2'(s) + a^2(\lambda_\ast\sin{\lambda_\ast s} + 2 \cos{\lambda_\ast s} - 2)^2 = 0.
\end{align}
Imposing $x_2(0)=0$ and integrating gives
\begin{subequations}
\begin{align}
    \lambda_0(s) &= C_{\lambda_0} + \frac{5a^2\lambda_\ast^2}{8} \left[\bigl(\lambda_\ast^2 - 4\bigr)\cos{2\lambda_\ast s} - 4\lambda_\ast\sin{2\lambda_\ast s}\right],\\
    x_2(s) &= \frac{a^2}{8\lambda_\ast} \left[\bigl(\lambda_\ast^2 - 4\bigr)\sin{2 \lambda_\ast s} + 32\sin{\lambda_\ast s} - 16\lambda_\ast \cos{\lambda_\ast s} + 4 \lambda_\ast\cos{2\lambda_\ast s}-2\lambda_\ast \bigl(\lambda_\ast^2s + 12s - 6\bigr)\right],
\end{align}
\end{subequations}
where, once again, $C_{\lambda_0} $ is a constant of integration not determined at this order. In now familiar fashion, requiring $x_2(1)=-d$ yields, once again, the relation \neqref{eq:a} between the buckling amplitude $a$ and $d$. At next order, we get
\begin{align}
    &\frac{a\lambda_\ast}{8}\left\{\lambda_\ast\cos{\lambda_\ast s} \left[-7a^2\lambda_\ast^2\bigl(\lambda_\ast^2 + 4\bigr) + 32b_{02}b_{20} - 32b_{11}^2(s - 1) + 16C_{\lambda_0}\right] + \lambda_\ast^2\left[2\sin{\lambda_\ast s}\bigl(7a^2\lambda_\ast^2 + 28a^2 - 8b_{11}^2s\bigr) \right.\right.\nonumber\\
    &\hspace{8mm}+ \left.24a^2\bigl(\lambda_\ast^2 - 4\bigr)\sin{2\lambda_\ast s} + 18a^2\bigl(4 - 3\lambda_\ast^2\bigr)\sin{3\lambda_\ast s} + 9a^2\lambda_\ast\bigl(\lambda_\ast^2 - 12\bigr)\cos{3\lambda_\ast s} + 96a^2\lambda_\ast\cos{2\lambda_\ast s}\right] \nonumber\\
    &\hspace{8mm}- \left.32\sin{\lambda_\ast s}\bigl(2b_{02}b_{20} + 2b_{11}^2 + C_{\lambda_0}\bigr) - 32\lambda_\ast\left[b_{02}b_{20}(2s - 1) + b_{11}^2 s\right]\right\} + \lambda_\ast^2y_3''(s) + y_3''''(s) = 0.
\end{align}
Together with $y_3(0)= y_3'(0)= y_3(1)= y_3'(1)= 0$, this yields
\begin{subequations}
\begin{align}
    y_3(s) &= -\frac{1}{192\lambda_\ast^3}\Bigl(16\sin{\lambda_\ast s} \left\{a^3\bigl(6s\lambda_\ast^4 - 3\lambda_\ast^4 - 48\lambda_\ast^2\bigr) + 2a\left[4b_{02}b_{20}\bigl(\lambda_\ast^2 - 2\lambda_\ast^2s + 16\bigr) + b_{11}^2\bigl(6\lambda_\ast^2s^2 + 2\lambda_\ast^2s - \lambda_\ast^2+23\bigr)\right]\right\}\nonumber\\ 
    &\hspace{18mm} + \lambda_\ast(2s - 1)\left[3a^3\lambda_\ast^2\bigl(\lambda_\ast^2 + 148\bigr) - 32a\lambda_\ast^2s\bigl(4b_{02}b_{20}s - 4b_{02}b_{20} + 2b_{11}^2s + b_{11}^2\bigr)- 48a\bigl(16b_{02}b_{20} + 9b_{11}^2\bigr)\right] \nonumber\\
    &\hspace{18mm}+ 192a^3\lambda_\ast^3\cos{2\lambda_\ast s} + 48a^3\lambda_\ast^2\bigl(\lambda_\ast^2 - 4\bigr)\sin{2\lambda_\ast s} + 6a^3\lambda_\ast^2\bigl(4 - 3\lambda_\ast^2\bigr)\sin{3\lambda_\ast s}+ 3a^3 \lambda_\ast^3\bigl(\lambda_\ast^2 - 12\bigr)\cos{3\lambda_\ast s}\nonumber\\
    &\hspace{18mm} + 16 a\lambda_\ast\cos{\lambda_\ast s}\left[2(2s + 3)\bigl(3a^2\lambda_\ast^2 - 8b_{02}b_{20}\bigr) - b_{11}^2\bigl(6\lambda_\ast^2s^2 - 8s + 27\bigr)\right]\Bigr)+C_{y_3}y_1^{\text{(iii)}}(s), \\ 
    C_{\lambda_0} &= \frac{1}{48}\left[3a^2\lambda_\ast^2\bigl(7\lambda_\ast^2 + 20\bigr) - 32b_{02}b_{20} + 8b_{11}^2\right].
\end{align}
\end{subequations}
The constant of integration $C_{y_3}$ does not need to be determined, because we can now compute the buckling forces at $s=0$ and $s=1$, finding
\begin{subequations}
\begin{align}
    F_0 &=\frac{\lambda_\ast^2}{\ell ^2} + \frac{1}{24}\left[3a^2 \lambda_\ast^2\bigl(\lambda_\ast^2 - 4\bigr) - 32b_{02}b_{20} + 8b_{11}^2\right] + O(\ell) = \frac{\lambda_\ast^2}{\ell ^2} + \frac{1}{6}\left[-2K + 3d\bigl(\lambda_\ast^2 - 4\bigr)\right] + O(\ell),\\
    F_1 &=F_0 + 2b_{20}^2\lambda_\ast^2 + O(\ell),
\end{align}
\end{subequations}
with the answer expressed again in terms of the Gaussian curvature $K= 4b_{02}b_{20}-b_{11}^2$ of the surface at the origin. As in the symmetric case, ``up'' and ``down'' buckling is equivalent, and the critical force is merely shifted compared to the flat case.

\subsection{Numerical solution of the buckling problem}
In this final section, we discuss the numerical solution of the Euler--Lagrange equations~\neqref{EL_complete}. We first note that these equations have an explicit integral
\begin{align}
\lambda(s)=\lambda_0+\dfrac{3\kappa(s)^2}{4},\label{eq:lsol}
\end{align}
where $\lambda_0$ is a constant of integration. This first integral is related to the tangential force balance on the elastic line~\cite{goldstein95}. A more pedestrian of deriving this is the following: Direct computation, facilitated by \textsc{Mathematica}, shows that
\begin{align}
0=2x'\left(L_{,x} - \frac{\mathrm{d}L_{,x'}}{\mathrm{d}s} + \frac{\mathrm{d}^2L_{,x''}}{\mathrm{d}s^2}\right)+2y'\left(L_{,y} - \frac{\mathrm{d}L_{,y'}}{\mathrm{d}s} + \frac{\mathrm{d}^2L_{,y''}}{\mathrm{d}s^2}\right)=4\lambda'(1-L_{,\lambda})-2\lambda\frac{\mathrm{d}L_{,\lambda}}{\mathrm{d}s}-\frac{\mathrm{d}^3L_{,\lambda}}{\mathrm{d}s^3}-6\kappa\kappa',
\end{align}
using the Euler--Lagrange equations~\eqref{EL_complete x} and \eqref{EL_complete y}. On imposing $L_{,\lambda}(s)=L_{,\lambda}'(s)=L_{,\lambda}'''(s)=0$ from \eqref{EL_complete lambda}, this immediately integrates to \eqref{eq:lsol}.

With this first integral, the Euler--Lagrange equations~\neqref{EL_complete x} and \neqref{EL_complete y} are fourth-order differential equations for $x(s),y(s)$, respectively, with one unknown parameter, $\lambda_0$. Eight of the nine boundary conditions needed are given by \eqsref{BC_relaxed} or \neqref{BC_buckling}. The final condition follows by differentiating \eqref{EL_complete lambda} to obtain $x'(s)x''(s)+y'(s)y''(s)=0$. For $s=0$, using $x'(0)=1$ and $y'(0)=0$, as required by \eqsref{BC_relaxed} or \neqref{BC_buckling}, this gives $x''(0)=0$, which is the final condition that we need.

\begin{acknowledgments}
This work was supported by the Max Planck Society. We thank J. Neipel and C. D. Modes for comments on the manuscript.
\end{acknowledgments}

\bibliography{arxiv_ref}

\end{document}